\documentclass[aps,prd,twocolumn,floatfix,nofootinbib,showpacs,superscriptaddress,tightenlines]{revtex4}

\usepackage{amsmath}
\usepackage{dcolumn}
\usepackage{lipsum}
\usepackage{amssymb}
\usepackage{epsfig}
\usepackage{graphicx}
\usepackage{amsmath}
\usepackage{bm}
\usepackage{setspace}
\usepackage{lscape}
\usepackage{amsthm}
\usepackage{bbold}
\usepackage{dcolumn}
\usepackage{epsfig}
\usepackage{graphics}
\usepackage{graphicx}
\usepackage{longtable}
\usepackage{color}
\usepackage{xcolor}
\usepackage{subfigure}
\usepackage{bm}
\usepackage{xspace}
\usepackage{cancel}
\usepackage{float}
\usepackage{multirow}
\definecolor{darkgreen}{rgb}{0,0.5,0}
\definecolor{purple}{rgb}{0.5,0,0.5}
\definecolor{nblue}{rgb}{0.0,0.0,0.50}
\definecolor{scarlet}{rgb}{1.0,0.2,0}
\definecolor{darkmagenta}{rgb}{0.55, 0.0, 0.55}
\definecolor{darkolivegreen}{rgb}{0.33, 0.42, 0.18}
\definecolor{darkcandyapplered}{rgb}{0.64, 0.0, 0.0}
\definecolor{linen}{rgb}{0.98, 0.94, 0.9}

\usepackage[colorlinks=true, pdfstartview=FitV, linkcolor=purple, citecolor= purple, urlcolor=blue]{hyperref}

\newcommand{\be}{\begin{equation}}
\newcommand{\tu}{\textcolor{red}{u}}

\newcommand{\fd}{\textcolor{blue}{f_1}}
\newcommand{\fdu}{\textcolor{blue}{f_2}}

\newcommand{\g}{\textcolor{black}{\Gamma}}
\newcommand{\D}{\textcolor{black}{\Delta}}
\newcommand{\td}{\textcolor{darkcandyapplered}{d}}
\newcommand{\tb}{\textcolor{blue}{b}}
\newcommand{\tc}{\textcolor{darkmagenta}{c}}
\newcommand{\ts}{\textcolor{darkgreen}{s}}
\newcommand{\ee}{\end{equation}}
\newcommand{\bea}{\begin{eqnarray}}
\newcommand{\eea}{\end{eqnarray}}
\newcommand{\beas}{\begin{eqnarray*}}
\newcommand{\eeas}{\end{eqnarray*}}
\newcommand{\nn}{\nonumber}

\newcommand{\tq}{\textcolor{red}{q}}
\newcommand{\tqu}{\textcolor{blue}{q_1}}

\newcommand{\MeV}{\text{MeV}} 
\newcommand{\GeV}{\text{GeV}} 
\newcommand{\rmh}{\hat{\alpha}_{\mathrm {IR}}}

\newcommand{\eqn}[1]{Eq.~(\ref{#1})}

\newcommand{\fig}[1]{Fig.~\ref{#1}}

\newcommand{\Meps}{\textcolor{blue}{{PS}}}
\newcommand{\Dps}{\textcolor{blue}{{DPS}}}
\newcommand{\Mv}{\textcolor{blue}{V}}
\newcommand{\Mav}{\textcolor{blue}{AV}}
\newcommand{\Ms}{\textcolor{blue}{S}}
\newcommand{\Ds}{\textcolor{blue}{{DS}}}
\newcommand{\Dv}{\textcolor{blue}{{DV}}}
\newcommand{\Dav}{\textcolor{blue}{{DAV}}}

\begin{document}
\title{First Radial Excitations of Baryons in a Contact Interaction: Mass Spectrum}
\author{L.X. Guti\'errez-Guerrero}
\email[]{lxgutierrez@conacyt.mx}
\affiliation{Instituto de F\'isica y Matem\'aticas, Universidad
Michoacana de San Nicol\'as de Hidalgo, Edificio C-3, Ciudad
Universitaria, Morelia, Michoac\'an 58040, M\'exico}
\affiliation{CONACyT-Mesoamerican Centre for Theoretical Physics,
Universidad Aut\'onoma de Chiapas, Carretera Zapata Km. 4, Real
del Bosque (Ter\'an), Tuxtla Guti\'errez 29040, Chiapas, M\'exico}
\author{Alfredo Raya}
\email[]{alfredo.raya@umich.mx}
\affiliation{Instituto de F\'isica y Matem\'aticas, Universidad
Michoacana de San Nicol\'as de Hidalgo, Edificio C-3, Ciudad
Universitaria, Morelia, Michoac\'an 58040, M\'exico}
\affiliation{Centro de Ciencias Exactas, Universidad del Bío-Bío,
Casilla 447, Chill\'an, Chile.}
\author{L. Albino}
\email[]{luis.albino.fernandez@gmail.com}
\affiliation{Instituto de F\'isica y Matem\'aticas, Universidad
Michoacana de San Nicol\'as de Hidalgo, Edificio C-3, Ciudad
Universitaria, Morelia, Michoac\'an 58040, M\'exico}
\affiliation{Dpto. Sistemas Físicos, Químicos y Naturales, Univ. Pablo de Olavide, 41013 Sevilla, Spain}
\author{R. J. Hern\'andez-Pinto}
\email[]{roger@uas.edu.mx}
\affiliation{Facultad de Ciencias F\'isico-Matem\'aticas, Universidad Aut\'onoma de Sinaloa, Ciudad Universitaria, Culiac\'an, Sinaloa 80000,
M\'exico}

\begin{abstract}
We compute masses of twenty positive parity first radial excitations of spin-$1/2$ and $3/2$ baryons composed of $\tu$, $\td$, $\ts$, $\tc$ and 
$\tb$ quarks in a quark-diquark picture within a contact interaction model. 
These excitations comprise of two elements: one characterized by a zero in the Faddeev amplitude, representing a radial excitation of the quark-diquark system and the other marked by a zero in the diquark's Bethe-Salpeter amplitude, corresponding to an intrinsic excitation of the diquark correlation. 
Wherever possible, we compare our results with other models and/or experiment.
We verify that the masses obtained through our model conform to the spacing rules for all the baryons studied, whether light or heavy and whether of spin 1/2 or 3/2. The computed masses do not just offer a guide to the future experimental searches but also compare well with the existing candidates for the possible radial excitations of some heavy baryons.  
\end{abstract}


\maketitle

\section{Introduction}
More than sixty years ago, the Faddeev equation was first proposed in Ref.~\cite{Faddeev:1960su} to investigate the three-body bound states. Therefore, this equation is ideally suited to study baryons. However, this problem can be reduced to two-body subsystems if we employ a quark-diquark picture. It is well-known by now that the difference between these two approaches yields a difference between the computed baryon masses of merely about 5\%, see Refs.~\cite{Eichmann:2011vu,Eichmann:2016yit}. The obvious advantage is that it considerably reduces mathematical and computational difficulties by converting a three-body problem into a two-body one. Gell-Mann suggested the idea of diquarks in his renowned article of 1964~\cite{Gell-Mann:1964ewy}. 
Within a couple of years, some of the earliest attempts to compute baryon masses were made in Refs.~\cite{Ida:1966ev,Lichtenberg:1967zz}. Static point-like diquarks 
were used in efforts to explain the problem of missing resonances as early as in 1969~\cite{Lichtenberg:1969sxc}.

The diquarks that we employ in our work are dynamical in nature with finite electromagnetic extent and an associated mass-scale which are both bounded below by the corresponding quantities characterizing the analogous mesonic system. 
A comprehensive review, Ref.~\cite{Barabanov:2020jvn}, and the references therein, are an excellent source on our improved understanding of diquarks  and the role they play in studying the baryon spectrum and their internal structure. Diquarks are paired color non-singlet correlations of two quarks. Owing to their color charge, these correlations are confined within baryons, tetra-quarks or penta-quarks, making direct observation impossible. 
Two quarks can form product representation in a color sextet or an antitriplet configuration. When exchanging a single gluon, a simple analysis of color flow indicates that  attractive interaction occurs between diquarks in a color antitriplet arrangement while it is repulsive for a sextet. Moreover, the diquark in a color antitriplet state can bind with a quark to create a color-singlet baryon. That is why these are dubbed as 
{\em good diquarks}~\cite{Wilczek:2004im}.
 Therefore, we shall exclusively focus on $qq$ states in a color antitriplet configuration. Similarly to mesons, various types of diquarks are distinguished by distinct Dirac structures. Therefore, we have scalar, pseudo-scalar, vector and axial-vector diquarks. 
 
 One expects dozens of baryonic 
states with singly, doubly, and triply heavy quarks. Studying radially excited baryons is naturally more complex than studying their counterpart ground states. Identification and analysis of these states is equally challenging in the experiments.
Radial excitations of baryons with two heavy quarks have been thoroughly studied through Poincar\'e-covariant Faddeev equation (FE) for baryons~\cite{Qin:2019hgk} (we refer to this model as QRS, after the initials of the authors of the Ref.~\cite{Qin:2019hgk}), relativistic quark model~\cite{Ebert:2002ig},
quark model~\cite{Yoshida:2015tia}, Salpeter model~\cite{Giannuzzi:2009gh}, hypercentral constituent quark model~\cite{Shah:2017liu,Shah:2016vmd}, the heavy quark effective theory (HQET)~\cite{Vishwakarma:2022vzy}, Faddeev method~\cite{Valcarce:2008dr}, QCD
sum rules~\cite{ShekariTousi:2024mso,Alrebdi:2022lat,Alomayrah:2020qyw} and  Regge phenomenology~\cite{Oudichhya:2021kop,Shah:2017jkr}.
These baryons are significant as they represent potential candidates for the states recently observed by the LHCb experiment. For example, initial observations of the baryon $\Omega_{\tb}^{-}$ with a mass of approximately 6.35 GeV suggest that this state corresponds to the first radial excitation of $\Omega_{\tb}^{-}(\ts\ts\tb)$ \cite{LHCb:2020tqd}. 
The LHCb
collaboration reported the observation of $\Xi_{\tb}^-$ with mass 6.226 GeV. From the observed mass and decay modes, it was suggested that $\Xi_{\tb}^-$ state might be $1P$ or $2S$ excited baryon~\cite {LHCb:2018vuc,Aliev:2018lcs}. 
LHCb has also detected two  excited $\Omega_{\tc}$ states decaying
to $\Xi_{\tc} K^-$ with masses of around 3.18 GeV and 3.3 GeV and which can be interpreted as radial excitations of $\Omega_{\tc}^0(\ts\ts\tc)$ and $\Omega_{\tc}^{0*}{(\ts\ts\tc)}$, respectively~\cite{Karliner:2023okv,LHCb:2023sxp,Pan:2023hwt}.
Contact interaction (CI) is a useful theoretical tool to explore the static properties of hadrons, particularly their masses. In Ref.~\cite{Paredes-Torres:2024mnz}, masses of radial excitations of mesons and diquarks were computed using this model. We extend that work to investigate baryons and compute the masses of their first radial excitations in the light and heavy sectors.
%
%
For that purpose, we have organized this paper as follows:
in Section \ref{Baryons}, a comprehensive examination of the FE for baryons with different spins is conducted; Section \ref{Excitations} is dedicated to elucidate the essential ingredients of our model to compute radial excitations rather accurately; Our results, along with the corresponding analysis, are detailed in Section~\ref{results}; and
finally, a summary and outlook is presented in Section~\ref{Conclusions}.
\section{Faddeev Equation }
\label{Baryons}
\begin{figure}[htpb]
\vspace{-3.0cm}
\hspace{-5mm}    \centerline{\includegraphics[scale=0.40,angle=-90]{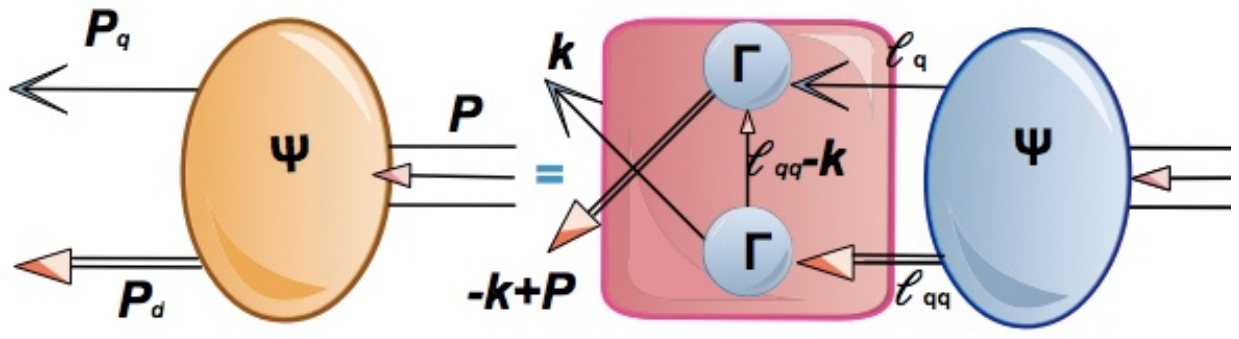}}       \vspace{-2.8cm}\caption{Poincar\'e covariant FE employed in this work to calculate masses of the first radial excitation of baryons. The red square represents the quark-diquark interaction Kernel. The single line denotes the dressed quark, the double line is the diquark while $\Gamma$ and $\Psi$ are the Bethe-Salpeter and Faddeev amplitudes (BSA and FA) , respectively. Configuration of momenta is: $\ell_{qq}=-\ell + P$, $k_{qq}=-k+P$, $P=P_d+P_q$.}
       \label{faddevv-Fig1}
\end{figure}

\begin{figure*}[htb]
\begin{center}
\caption{\label{octydec}\textit{Left}: \textbf{Baryons with spin $1/2$}.- We show the mixed-symmetric $20$-plet. Note that all the ground-state baryons in this multiplet have $J^P =(1/2)^+$. It has the $SU(3)$ octet on the lowest layer. The singly heavy baryons are composed of a bottom quark and two light quarks $(\tu,\td,\ts)$, located in the second layer. The doubly heavy baryons are positioned in the top-most layer.
\textit{Right}: \textbf{Baryons with spin $3/2$}.- It depicts states of baryons with spin $3/2$ made from four quarks of the types $\tu, \td, \ts$, and $\tb$. Doubly heavy baryons and triply heavy baryons are localized in the highest layers.}\label{figures} \label{figura3}
\vspace{-3cm}
\begin{minipage}[c]{8cm}
\includegraphics[width=10cm]{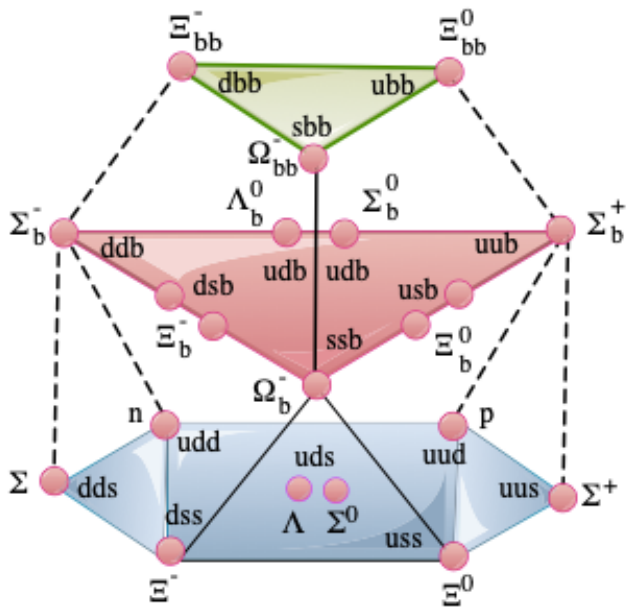}
\end{minipage}
\vspace{-3.5cm}
\begin{minipage}[c]{8cm}
\includegraphics[width=10cm]{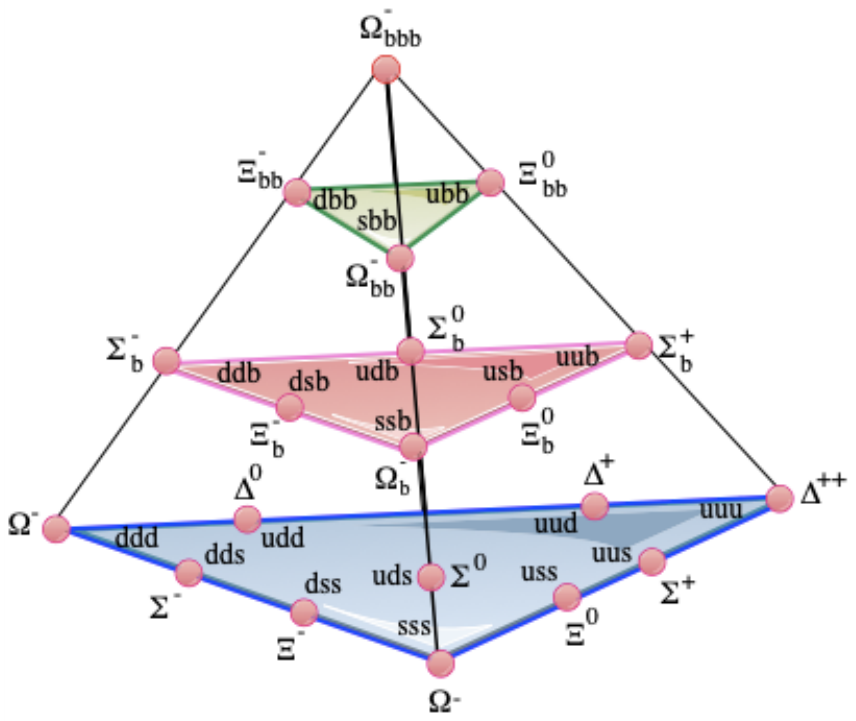}
\end{minipage}
\vspace{-0cm}
\end{center}
\end{figure*}

%
We base our description of baryons-bound states in the quark-diquark approximation on FE, which is illustrated in Fig.~\ref{faddevv-Fig1}.
\\
In this section, we shall examine this FE in detail and the requisite components for our computation, adhering closely to the notation employed in Ref.~\cite{Gutierrez-Guerrero:2019uwa}.
Using this equation, we can predict the masses of ground and radially excited baryons with $J^P = 1/2^+$ and $J^P = 3/2^+$, where $J$ is the total spin and $P$ denotes the parity. \\
Fig.~\ref{octydec} presents
the $SU(5)$ flavor depiction of these states. The baryons multiplets that arise from $3\otimes3\otimes3$ are: a decouplet, two octets and a singlet.
The corresponding multiplet structure for $SU(4)$ is $4\otimes4\otimes 4 = 20_S\oplus20_{MS}\oplus20_{MA}\oplus 4_A$. 
Note that explicit quark masses break the flavor symmetry. The larger the group, the more significant the amount
of breaking as each new quark is significantly heavier. However, this group algebra approach helps us identify the baryons whose masses we are interested to compute. 
As only a representative example, we present such baryon multiplets with $\tu$, $\td$, $\ts$ and $\tb$ quark in \fig{octydec}. The multiplet with charm quarks is analogous to the one containing the bottom quark.
A detailed examination of the FE delineating states both with $J=1/2$ and $J=3/2$ will be undertaken in the ensuing subsections.
\subsection{Baryons with spin 1/2}
\label{baryons-1}
The mass of the ground-state baryon with spin $1/2$ comprised by the quarks $[\tq\tq\tqu]$ and the  momentum configuration given in \fig{faddevv-Fig1} is determined by a $5\times 5$ matrix FE.
It can be written in the following form~:
\begin{eqnarray}
\nonumber
\lefteqn{
 \left[ \begin{array}{r}
{\cal S}(k;P)\, u(P)\\
{\cal A}^i_\mu(k;P)\, u(P)
\end{array}\right]}\\
& =&  -\,4\,\int\frac{d^4\ell}{(2\pi)^4}\,{\cal M}(k,\ell;P)
\left[
\begin{array}{r}
{\cal S}(\ell;P)\, u(P)\\
{\cal A}^j_\nu(\ell;P)\, u(P)
\end{array}\right] .\rule{1em}{0ex}
\label{FEone}
\end{eqnarray}
The general matrices ${\cal S}(\ell;P)$ and ${\cal A}^i_\nu(\ell;P)$, which describe the momentum-space correlation between a quark and a diquark in the nucleon and the roper are described in Refs.~\cite{Oettel:1998bk, Cloet:2007pi}.  However, with the CI employed in this article, they simplify considerably~:
\begin{subequations}
\label{FaddeevAmp}
\begin{eqnarray}
{\cal S}(P) &=& s(P) \,\mbox{\boldmath $I$}_{\rm D}\,,\\
{\cal A}^i_\mu(P) &=& a_1^i(P) \gamma_5\gamma_\mu + a_2^i(P) \gamma_5 \hat P_\mu \,, \quad i=+,0\rule{2em}{0ex}
\end{eqnarray}
\end{subequations}
where the scalars $s$ and $a_{1,2}^i$ are independent of the relative quark-diquark momentum and $\hat P^2=-1$. FA is thus represented by the eigenvector~:
\begin{equation}
\Psi(P) = \left[\begin{array}{c}
s(P)\\[0.7ex]
a_1^+(P)\\[0.7ex]
a_1^0(P)\\[0.7ex]
a_2^+(P)\\[0.7ex]
a_2^0(P)\end{array}\right] \,.
\end{equation}
The Kernel in Eq.~(\ref{FEone}) can generically be expanded as
\begin{eqnarray}
\label{calM} {\cal M}(k,\ell;P) =
 \begin{pmatrix} \mathcal{M}^{11}&\mathcal{M}^{12}_\nu&\mathcal{M}^{13}_\nu&\mathcal{M}^{14}_\nu&\mathcal{M}^{15}_\nu\\ \\
 \mathcal{M}^{21}_\mu& \mathcal{M}^{22}_{\mu\nu}&\mathcal{M}^{23}_{\mu\nu}& \mathcal{M}^{24}_{\mu\nu}&\mathcal{M}^{25}_{\mu\nu}\\ \\
 \mathcal{M}^{31}_{\mu}&\mathcal{M}^{32}_{\mu\nu}& \mathcal{M}^{33}_{\mu\nu}&\mathcal{M}^{34}_{\mu\nu}&\mathcal{M}^{35}_{\mu\nu}\\ \\
 \mathcal{M}^{41}_\mu& \mathcal{M}^{42}_{\mu\nu}&\mathcal{M}^{43}_{\mu\nu}&\mathcal{M}^{44}_{\mu\nu}&\mathcal{M}^{45}_{\mu\nu}\\ \\
\mathcal{M}^{51}_{\mu}&\mathcal{M}^{52}_{\mu\nu}& \mathcal{M}^{53}_{\mu\nu}&\mathcal{M}^{54}_{\mu\nu}&\mathcal{M}^{55}_{\mu\nu} \\
 \label{B-matrix} \end{pmatrix} \,. \end{eqnarray}
The elements of the matrix in Eq.~(\ref{B-matrix}) are detailed in Appendix~\ref{app:Fad}. In order to simplify Eqs.~(\ref{FEone}), we use static approximation for the exchanged quark with mass $M$ and flavor $f$. It was introduced first in Ref.~\cite{Buck:1992wz}~:
\bea
S(p)=\frac{1}{i \gamma \cdot p + M_f} \to
\frac{1}{M_f} \,. \eea
A variation of it was implemented in Ref.~\cite{Xu:2015kta},
\bea
S(p)=\frac{1}{i \gamma \cdot p + M_f}\to
\frac{g_{B}}{i \gamma \cdot p + M_f}  \,. \eea
We follow Refs.~\cite{Roberts:2011cf,Lu:2017cln,Chen:2012qr} and represent the quark (propagator) exchanged between the diquarks as
\begin{equation}
S^{\rm T}(k) \to \frac{g_B}{M_f}\,.
\label{staticexchangec}
\end{equation}
The superscript ``T'' indicates matrix transpose.
However, in the implementation of this framework to heavy baryons 
in the ground state with spin
$1/2$, it suffices to use $g_B = 1$, as discussed in reference \cite{Gutierrez-Guerrero:2019uwa}. For radial excitations, however, this
value will no longer be $1$, as will be explained in the subsequent sections.\\
The spin-$1/2$ heavy baryons are represented by the
following column matrices:
 \begin{equation}
\nn \begin{array}{cc}
u_{\Xi_{cc}^{++} (\tu\tc\tc)}= \left[
\begin{array}{c}
[\tu\tc] \tc \\
\{\tc\tc\}\tu\\
\{\tu\tc\}\tc\\
\end{array} \right], &
u_{\Omega_{\tc\tc}^+(\ts\tc\tc)}=
\left[ \begin{array}{c}
[\ts\tc] \tc \\
\{\tc\tc\} \ts \\
\{\ts\tc\} \tc \\
\end{array} \right],
\end{array} 
\end{equation}

 \begin{equation}
\nn \begin{array}{cc}
u_{\Omega_{\tc}^0(\ts\ts\tc)}= \left[
\begin{array}{c}
[\ts\tc] \ts \\
\{\ts\ts\}\tc\\
\{\ts\tc\}\ts\\
\end{array} \right], &
u_{\Sigma_{\tc}^{++}(\tu\tu\tc)}=
\left[ \begin{array}{c}
[\tu\tc] \tu \\
\{\tu\tu\} \tc \\
\{\tu\tc\} \tu \\
\end{array} \right],
\end{array} 
\end{equation}

 \begin{equation}
\nn \begin{array}{cc}
u_{\Xi_{bb}^0(\tu\tb\tb)}= \left[
\begin{array}{c}
[\tu\tb] \tb \\
\{\tb\tb\}\tu\\
\{\tu\tb\}\tb\\
\end{array} \right], &
u_{\Omega_{bb}^-(\ts\tb\tb)}=
\left[ \begin{array}{c}
[\ts\tb] \tb \\
\{\tb\tb\} \ts \\
\{\ts\tb\} \tb \\
\end{array} \right],
\end{array} 
\end{equation}
 \begin{equation}
\nn \begin{array}{cc}
u_{\Omega_b^{-}(\ts\ts\tb)}= \left[
\begin{array}{c}
[\ts\tb] \ts \\
\{\ts\ts\}\tb\\
\{\ts\tb\}\ts\\
\end{array} \right], &
u_{\Sigma_b^+(\tu\tu\tb)}=
\left[ \begin{array}{c}
[\tu\tb] \tu \\
\{\tu\tu\} \tb \\
\{\tu\tb\} \tu \\
\end{array} \right],
\end{array} 
\end{equation}
 \begin{equation}
\nn \begin{array}{cc}
u_{\Omega(\tc\tc\tb)}= \left[
\begin{array}{c}
[\tc\tb] \tc \\
\{\tc\tc\}\tb\\
\{\tc\tb\}\tc\\
\end{array} \right], &
u_{\Omega(\tc\tb\tb)}=
\left[ \begin{array}{c}
[\tc\tb] \tb \\
\{\tb\tb\} \tc \\
\{\tc\tb\} \tb \\
\end{array} \right].
\end{array} 
\end{equation}
Note that we use the notation introduced in Ref.~\cite{Chen:2012qr}, [$\tq\tqu$] for scalar diquarks, and $\{\tq\tq\},\{\tq\tqu\}$ for axial-vector diquarks. 
The Faddeev amplitude for baryons with spin $3/2$ is different from that for spin $1/2$.
Below we discuss the fundamentals of the FE and FA of these baryons.
 %
\subsection{Baryons with spin-3/2}
\label{baryons-2}
Baryons with spin-$3/2$ are specially important because they can involve states with three $\tc$-quarks and three $\tb$-quarks. In order to calculate their masses, we note that it is not possible to combine a spin-zero
diquark with a spin-1/2 quark to obtain spin-$3/2$ baryon. Hence, such a baryon is comprised solely of axial-vector correlations.
The FA for the positive-energy baryon
is~: 
\bea
\Psi_\mu = \psi_{\mu\nu}(P)u_\nu(P) \;, \nn 
\eea
where $P$ is the total momentum of the baryon, $u_\nu(P)$ is a Rarita-Schwinger spinor,
\bea \label{fd1}
\psi_{\mu\nu}(P)u_{\nu} &=& \Gamma_{\tq\tqu, \mu} \Delta_{\mu\nu,_{\tq\tq}}^{1^+}(\ell_{\tq\tq}){\cal D}_{\nu\rho}(P)u_{\rho}(P) \,,
\eea
with $\Delta_{\mu\nu,\tq\tq}^{1^+}$ is the diquark propagator defined in Appendix~\ref{app:Fad} and,
\bea
\label{DeltaFA}
{\cal D}_{\nu\rho}(\ell;P) &=& {\cal S}(\ell;P) \, \delta_{\nu\rho} + \gamma_5{\cal A}_\nu(\ell;P) \,\ell^\perp_\rho \,.
\eea
Understanding the structure of these states is simpler than in the case of the nucleon.  We have provided more details in Appendix~\ref{App:EM}.
%
In what follows, we only consider the baryons with two possible structures: $\tq\tq\tq$ and $\tqu\tq\tq$. \\ \\
%
\noindent
{\bf Baryons ($\tq\tq\tq$):} Only one possible type of diquarks exists for a baryon composed of same three quarks $(\tq\tq\tq)$. In this case, the FA is:
\begin{equation}
{\cal D}_{\nu\rho}(\ell;P) u_\rho^B(P) = f^B(P) \, \mathbf{I}_{\rm D} \, u_\nu^B(P)\,.
\label{DnurhoI}
\end{equation}
Employing Feynman rules for Fig.~\ref{faddevv-Fig1} and using the expression for the FA, Eq.~({\ref{DnurhoI}}), we can write
\bea
f^B(P) u_\mu^B(P)=
4\frac{g_B}{M_{\tq}}\int \frac{d^4 \ell}{(2\pi)^4}\mathcal{M_{\mu\nu}}f^B(P) u_\nu^B(P) \,,
\eea
where we have suppressed the functional dependence of $\mathcal{M}_{\mu\nu}$ on 
momenta for simplicity, and
\bea
\nn\mathcal{M}_{\mu\nu} &=&2i\g^{1^+}_{\{\tq\tq\},\rho}(l_{\tq\tq})i\overline{\g}^{1^+}_{\{\tq\tq\},\mu}(-k_{\tq\tq})S_{\tqu}(l_{\tqu})\D^{1^+}_{\{\tq\tq\},\rho\nu}(l_{\tq\tq})\;.
\eea
We 
now multiply both sides by ${\bar{u}}^B_{\beta}(P)$ from the left and sum over the polarization 
not explicitly shown here, to obtain
\bea
\Lambda_{+}(P)R_{\mu\beta}(P)=4\frac{g_B}{M_{\tq}}\int \frac{d^4 \ell}{(2\pi)^4}\mathcal{M_{\mu\nu}}\Lambda_{+}(P)R_{\nu\beta} \,.
\eea
 Positive energy projection operator $\Lambda_+$ and  $R_{\mu\beta}$ are defined in Eqs.~(\ref{Lplus}) and (\ref{Rde}) in Appendix~\ref{App:EM}.
Finally we contract with $\delta_{\mu\beta}$ and with the aid of a Feynman parametrisation, we obtain
\begin{eqnarray}\nn
2\pi^2 = \frac{g_B}{M_{\tq}}\frac{E_{\{\tq\tq\}_{1^+}}^2}{m_{\{\tq\tq\}_{1^+}}^2}
 \hspace{-2mm} \int_0^1 \hspace{-2mm} d\alpha\, {\cal L}^\Omega \overline{\cal C}^{\rm iu}_1(\omega (\alpha,M_{\tq}^2,m^2_{\{\tq\tq\}_{1^+}},P^2))\,,\\
 \label{ef}
\end{eqnarray}
%
where $E_{\{\tq\tq\}_{1^+}}$ is the structure in the BSA in \eqn{BSA-diquarks} in Appendix~\ref{BS-s}. The function ${\cal C}^{\rm iu}_1$ is defined in \eqn{Ciu} in the Appendix~ \ref{CI-1}, and $m_{\{\tq\tq\}_{1^+}}$ is de diquark mass. We have defined
\bea 
\hspace{-1cm}\nn {\cal L}^\Omega &=& [m_{\{\tq\tq\}_{1^+}}^2 - (1-\alpha)^2 P^2][-\alpha P^2 + M_{\tq}]
\,,
\eea
\bea
\omega (\alpha,M_{\tq}^2,m^2_{\{\tq\tq\}_{1^+}},P^2)\nn &=&M_{\tq}^2 (1-\alpha) + \alpha m^2_{\{\tq\tq\}_{1^+}}  \\ \nn 
\hspace{1cm}&+& \alpha(1-\alpha) P^2\,,
\end{eqnarray}
The color-singlet bound states constructed from three identical heavy charm/bottom quarks are:
\begin{eqnarray}
&&\nn \begin{array}{cc}
u_{\Omega_{\tc\tc\tc}^{++*}}= \left[
\begin{array}{c}
\{\tc\tc\} \tc \\
\end{array} \right], &
\hspace{1.0cm}u_{\Omega_{\tb\tb\tb}^{-*}}=
\left[ \begin{array}{c}
\{\tb\tb\} \tb \\
\end{array} \right],
\end{array}
\end{eqnarray}
From the \eqn{ef}, it is straightforward to compute the mass of these baryons.\\ \\
\noindent
{\bf Baryons($\tqu\tq\tq$):}
 For a baryon with the quark component structure ($\tqu\tq\tq$), there are two possible diquarks, $\{\tq\tq\}$ and $\{\tqu\tq\}$. The FA for such a baryon is:
\begin{equation}  \label{d-diquark}
{\cal D}_{\nu\mu}^{B}(P)u_{\mu}^{B}(P;s) =
\sum_{i=\{\tqu\tq\},\{\tq\tq\}} d^{i}(P) \delta_{\nu\lambda} u_{\lambda}^{B}(P;s),
\end{equation}
so that the corresponding FE has the form
\begin{equation}
\left[\begin{matrix}
d^{\{\tqu\tq\}}\\ d^{\{\tq\tq\}}\end{matrix}\right]
u_{\mu}^{B}= -4 \int \frac{d^{4}l}{(2\pi)^{4}} \\
{\cal M}
\left[\begin{matrix}
d^{\{\tqu\tq\}}\\ d^{\{\tq\tq\}}\end{matrix}\right] u_{\nu}^{B} \,,
\label{eq:D2}
\end{equation}%
where
\bea
\label{M32}
\cal{M}=\left[\begin{matrix}
{\cal M}_{\mu\nu}^{11} & {\cal M}_{\mu\nu}^{12} \\
\rule{0ex}{3.5ex}
{\cal M}_{\mu\nu}^{21} & {\cal M}_{\mu\nu}^{22}
\end{matrix}\right]\;,
\eea
with the elements of the matrix ${\cal M}$ given by~:
\begin{equation} 
\begin{split}
{\cal M}_{\mu\nu}^{11}& =t^{f_{11}}\frac{\textcolor{red}{g_B}}{M_{\tqu}} \,
\Gamma_{\rho}^{1^{+}}(\ell_{\tqu\tq})  \,
\bar{\Gamma}_{\mu}^{1^{+}}(-k_{\tqu\tq}) \, S(l_{\tq}) \,
\Delta_{\rho\nu}^{1^{+}}(\ell_{\tqu\tq}), \\
{\cal M}_{\mu\nu}^{12}&=t^{f_{12}}\frac{\textcolor{red}{g_B}}{M_{\tq}} \,
\Gamma_{\rho}^{1^{+}}(\ell_{\tq\tq})  \,
\bar{\Gamma}_{\mu}^{1^{+}}(-k_{\tqu\tq}) \, S(l_{q_1}) \,
\Delta_{\rho\nu,\{\tq\tq\}}^{1^{+}}(\ell_{\tq\tq}),\\
%
{\cal M}_{\mu\nu}^{21} &=t^{f_{21}}\frac{\textcolor{red}{g_B}}{M_{\tq}} \,
\Gamma_{\rho}^{1^{+}}(\ell_{\tqu\tq})  \,
\bar{\Gamma}_{\mu}^{1^{+}}(-k_{\tq\tq}) \, S(l_{\tq}) \,
\Delta_{\rho\nu}^{1^{+}}(\ell_{\tqu\tq}), \\
{\cal M}_{\mu\nu}^{22}&=t^{f_{22}} \frac{\textcolor{red}{g_B}}{M_{\tq}} \,
\Gamma_{\rho}^{1^{+}}(\ell_{\tq\tq})  \,
\bar{\Gamma}_{\mu}^{1^{+}}(-k_{\tq\tq}) \, S(l_{\tq_1}) \,
\Delta_{\rho\nu}^{1^{+}}(\ell_{\tq\tq}), 
\end{split}
\end{equation}
where $t^f$ are the flavor matrices and can be found in Appendix~\ref{app:Fla}. The color-singlet bound states constructed from three heavy charm/bottom quarks are:
\begin{eqnarray}
&&\nn \begin{array}{cc}
u_{\Omega_{\tc\tc\tb}^{+*}}= \left[
\begin{array}{c}
\{\tc\tc\}\tb \\
\{\tc\tb\} \tc \\
\end{array} \right], &
\hspace{1cm}u_{\Omega_{\tc\tb\tb}^{0*}}=
\left[ \begin{array}{c}
\{\tc\tb\} \tb \\
\{\tb\tb\} \tc
\end{array} \right].
\end{array} 
\end{eqnarray}
The column vectors representing singly and doubly heavy baryons are:
\begin{eqnarray}
\hspace{-0.8cm}&&\nn \begin{array}{cc}
u_{\Sigma_{\tc}^{++*}{(\tu\tu\tc)}}=
\left[ \begin{array}{c}
\{\tu\tu\} \tc \\
\{\tu\tc\} \tu
\end{array} \right],
&
\hspace{0.3cm}u_{\Xi_{\tc\tc}^{++*}{(\tu\tc\tc)}}= \left[
\begin{array}{c}
\{\tu\tc\} \tc \\
\{\tc\tc\} \tu \\
\end{array} \right], 
\end{array} \\ \nn \\
\hspace{-0.8cm}&&\nn \begin{array}{cc}
u_{\Omega_{\tc}^{0*}{(\ts\ts\tc)}}=
\left[ \begin{array}{c}
\{\ts\ts\} \tc \\
\{\ts\tc\} \ts
\end{array} \right],
 &
\hspace{0.9cm}u_{\Omega_{\tc\tc}^{+*}{(\ts\tc\tc)}}= \left[
\begin{array}{c}
\{\ts\tc\} \tc \\
\{\tc\tc\} \ts \\
\end{array} \right], 
\end{array}
\\ \nn \\
\hspace{-0.8cm}&&\nn \begin{array}{cc}
u_{\Sigma_{\tb}^{+*}{(\tu\tu\tb)}}=
\left[ \begin{array}{c}
\{\tu\tu\} \tb \\
\{\tu\tb\} \tu
\end{array} \right],
&
\hspace{0.7cm}u_{\Sigma_{\tb\tb}^{0*}{(\tu\tb\tb)}}= \left[
\begin{array}{c}
\{\tu\tb\}\tb \\
\{\tb\tb\} \tu \\
\end{array} \right],
\end{array} \\ \nn \\
\hspace{-0.8cm}&&\nn \begin{array}{cc}
u_{\Omega_{\tb}^{-*}{(\ts\ts\tb)}}=
\left[ \begin{array}{c}
\{\ts\ts\} \tb \\
\{\ts\tb\} \ts
\end{array} \right],
&
\hspace{0.8cm}u_{\Omega_{\tb\tb}^{-*}{(\ts\tb\tb)}}= \left[
\begin{array}{c}
\{\ts\tb\}\tb \\
\{\tb\tb\} \ts \\
\end{array} \right] ~.
\end{array} 
\end{eqnarray}
Now that we have introduced all the essential elements to study baryons through the FE, we can move on to discuss the crucial difference between studying ground and excited states in the next Section.

\section{First Radial Excitations of Baryons}
\label{Excitations}
For the sake of comparison with other works, experimental results and PDG, we use the spectroscopic notation $n\,^{2S+1}L_J$, where $J$ is the total angular momentum, $S$ is the spin, $n$ is the principal quantum number, and $L$ is the orbital angular momentum. 
The ground state is represented by $n=0$ in this notation and the first radial excitation by $n=1$. 
The spectroscopic notation for the baryons in the first radial excitation is, therefore, 
\bea
\nn \text{Spin }1/2 &\to& 1 {~}^2S_{1/2}\,, \\
\text{Spin }3/2 &\to& 1 {~}^4S_{3/2} \,.
\eea
Let us now recall that the leading Chebyshev moment of the FA for the first radial excitation of a baryon should have a single zero, similar to mesons~\cite{Holl:2004fr}.
However, note that the scenario concerning the radial excitations of baryons is more intricate than that of mesons or diquarks. In addition to accounting for the zero in the FA, one must also consider that the diquarks themselves might be radially excited. 

To obtain the mass and amplitude associated
with the first radial excitation of a diquark comprised of a quark with
flavor $\fd$ and another with flavor $\fdu$,
 we employ the same methods as detailed in Refs.~\cite{Gutierrez-Guerrero:2019uwa,Gutierrez-Guerrero:2021rsx,Paredes-Torres:2024mnz}. In fact, for our calculations, we incorporate the masses of the excited diquarks in Table~\ref{fbod} as were computed in~\cite{Paredes-Torres:2024mnz} employing the CI.
\begin{table*}[htbp] 
    \centering
    \caption{\label{fbod} Scalar and axial-vector diquark masses in its first radial excitation obtained in \cite{Paredes-Torres:2024mnz}
        using the parameters described in Appendix \ref{CI-1}. 
        In calculating the mass of radial excitations of diquarks, the value of $d_F$ is derived from an equation obtained through a fit to the masses of meson excitations. For further details, we refer the reader to Ref.~\cite{Paredes-Torres:2024mnz}.}
    \renewcommand{\arraystretch}{1.6} %
    \begin{tabular}{@{\extracolsep{0.5 cm}}c|cccccccccc}
         \hline
        \hline
Scalar Diquark        & $\tu\td$ & $\tu\ts$ & $\ts\ts$ & $\tc\tu$ & $\tc\ts$ & $\tu\tb $ & $\ts\tb$ & $\tc\tc$ & $\tc\tb$ & $\tb\tb$
         \\  
         & 1.28 & 1.52 &1.72 &2.53 & 2.78 &5.68 & 5.94 & 3.90 & 6.80 &9.68\\
Axial-vector Diquark        & $\tu\td$ & $\tu\ts$ & $\ts\ts$ & $\tc\tu$ & $\tc\ts$ & $\tu\tb $ & $\ts\tb$ & $\tc\tc$ & $\tc\tb$ & $\tb\tb$\\
        & 1.48 & 1.63 & 1.80 & 2.57 & 2.78 & 5.68 & 5.91 & 3.92 & 6.85 & 9.71\\
        \hline
        \hline
    \end{tabular}
    \end{table*}
 However, we naturally include an extra term associated with the first radial excitation possessing a single
zero, just like the radial wave function for bound states within any sophisticated QCD-based treatment of mesons.
In the phenomenological application of CI, we follow the methodology in Refs.~\cite{Roberts:2011cf,Roberts:2011wy} and insert a zero by hand into the kernels in the FE, multiplying it by $(1-q^2d_F)$
which forces a zero into the kernel at $q^2 = 1/d_F$ , where $d_F$ is an additional parameter. The presence of this zero
reduces the coupling in the FE and increases the mass of the bound-state. The presence of this term modifies the functions ${\cal C}^{\rm iu}$ in Ref.~\cite{Gutierrez-Guerrero:2021rsx} and now it must be replaced by ${\cal F}^{\rm iu} = {\cal C}^{\rm iu} - d_F {\cal D}^{\rm iu}$ where
\begin{eqnarray}
\nonumber
{\cal D}^{\rm iu}(\omega) = \int_0^\infty ds\,s\,\frac{s}{s+\omega }
\to  \int_{\tau_{\rm uv}^2}^{\tau_{\rm iu}^2} d\tau\, \frac{2}{\tau^3} \,
\exp\left[-\tau \omega\right], \rule{1em}{0ex}
\end{eqnarray}
${\cal F}^{\rm iu}_1(z) = - z (d/dz){\cal F}^{\rm iu}(z)$ and $\overline{\cal F}_1(z) = {\cal F}_1(z)/z$.  The general decomposition of the bound states FE for radial excitations is the same as the ground state. 
Now, we only need to discuss how to choose the location of the zero in the excitation. For this purpose, it is necessary to fix the parameter $d_F$; this value was set to $1/d_F=2M^2$ in Refs.~\cite{Roberts:2011cf,Roberts:2011wy} 
for calculating radial excitations of light mesons in pseudoscalar and vector channels. However, this value was selected as Ref.~\cite{Paredes-Torres:2024mnz} for heavy and heavy-light mesons and diquarks. Nevertheless, the resulting BSA from this $d_F$ selection proved exceedingly minute for diquarks incorporating heavy quarks. This tendency is particularly conspicuous in the scenario of heavy-light compositions. These values inevitably impact our baryon mass calculations, as we necessitate precise knowledge of the implicated amplitudes and masses for diquarks. This observation compels us to carefully choose our parameters so that the amplitudes yield minimal influence over our calculations.
In the case of radial excitations of light baryons using the CI model, the value of $d_F$ had been utilized as $1/d_F=2M^2(1\pm 0.2)$.
We have modified and unified this value in the case of baryons containing heavy quarks. For both baryons with spin-1/2 and 3/2, we use the value $d_F = 0.1\,\text{GeV}^{-2}$.\\
In reference \cite{Paredes-Torres:2024mnz}, a more sophisticated method was used to obtain $d_F$; however, once the diquark masses are determined using this mechanism, obtaining this parameter for baryons becomes straightforward. Our immediate benefit is the reduction in the number of parameters used in our model.
On the other hand, another parameter we need to fix for our calculations is the value of $g_B$ in Eq.~(\ref{staticexchangec}). For light baryons in the ground state using CI, we set $g_B = 1.18$ for the baryon octet and $g_B =1.56$ for the baryon decuplet \cite{Roberts:2011cf,Chen:2012qr}. Therefore, it modified in \cite{Gutierrez-Guerrero:2019uwa, Gutierrez-Guerrero:2021rsx} for heavy and heavy-light baryons in the ground state, where $g_B = 0.75$ was employed for baryons with spin-1/2, and $g_B = 1$ for baryons with spin-3/2.
For the first radial baryon excitations, the selection of $g_B$ values adheres to the following criteria~:
\\\\
\textbf{\underline{Baryons spin-$1/2$~:}} Deriving the mass of these baryons involves the scalar and axial-vector diquark amplitudes, such that
\bea
\nn g_B E_{[\tq\tqu]{0^+}}E_{[\tq\tqu]{0^+}}&\to& g_B\,,\\
\nn g_B E_{[\tq\tqu]{0^+}}E_{\{\tq\tq\}{1^+}} &\to& g_B\,,\\
\nn g_B E_{[\tq\tqu]{0^+}}E_{\{\tq\tqu\}{1^+}} &\to& g_B\,,\\
\nn g_B E_{\{\tq\tq\}{1^+}}E_{\{\tq\tq\}{1^+}} &\to& g_B\,,\\
\label{gbs} g_B E_{\{\tq\tqu\}{1^+}}E_{\{\tq\tqu\}{1^+}} &\to& g_B\,.
\eea
For baryons composed of one or no heavy quarks with spin-1/2, we use the value $g_B=3.5$. However, for baryons with double and triple heavy quarks, we use $g_B=1$.
\\\\
\textbf{\underline {Baryons spin-$3/2$:}} 
In this case, the baryon is composed only of axial-vector diquarks; then we have two cases: Baryons with $(\tq\tq\tqu)$ or baryons with three identical quarks ($\tq\tq\tq$).
For different quarks:
\bea
g_B E_{\{\tq\tqu\}{1^+}}E_{\{\tq\tq\}{1^+}} \to g_B\,.
\eea
When the three quarks are identical, as in the case of the heaviest baryons
\bea
g_B E_{\{\tq\tq\}{1^+}}E_{\{\tq\tq\}{1^+}} \to g_B\,.
\eea
For this case, for baryons composed of one or no heavy quarks, we use 
$g_B=10$ and for double and triple heavy baryons, we use $g_B=7$.\\
Finally, employing the diquark masses in Table~\ref{fbod}  obtained in Ref.~\cite{Paredes-Torres:2024mnz} for radial excitations, 
the parameters described above and Tables \ref{parameters1} and \ref{table-M} in Appendix \ref{CI-1}, we present our results in the next section alongside those from other models.
\section{Results}
\label{results}
This Section unveils our numerical findings concerning the baryon masses for the states $n=0$ and $n=1$. To enhance the depth of our analysis, we employ graphical illustrations, providing a visual narrative of our work alongside comparisons with alternative models. 
In Table \ref{fbod}, we show the masses of the excited scalar and axial-vector diquarks obtained in the article, which are necessary for calculating the masses of excited baryons.
\subsection{First Radial Excitation: Baryons with spin-1/2}
\begin{figure}[htpb]
\caption{\label{oct-lig}  Differences between the ground states and their first radial excitations for light baryons}
       \vspace{-0cm}
       \centerline{\includegraphics[scale=0.23,angle=0]{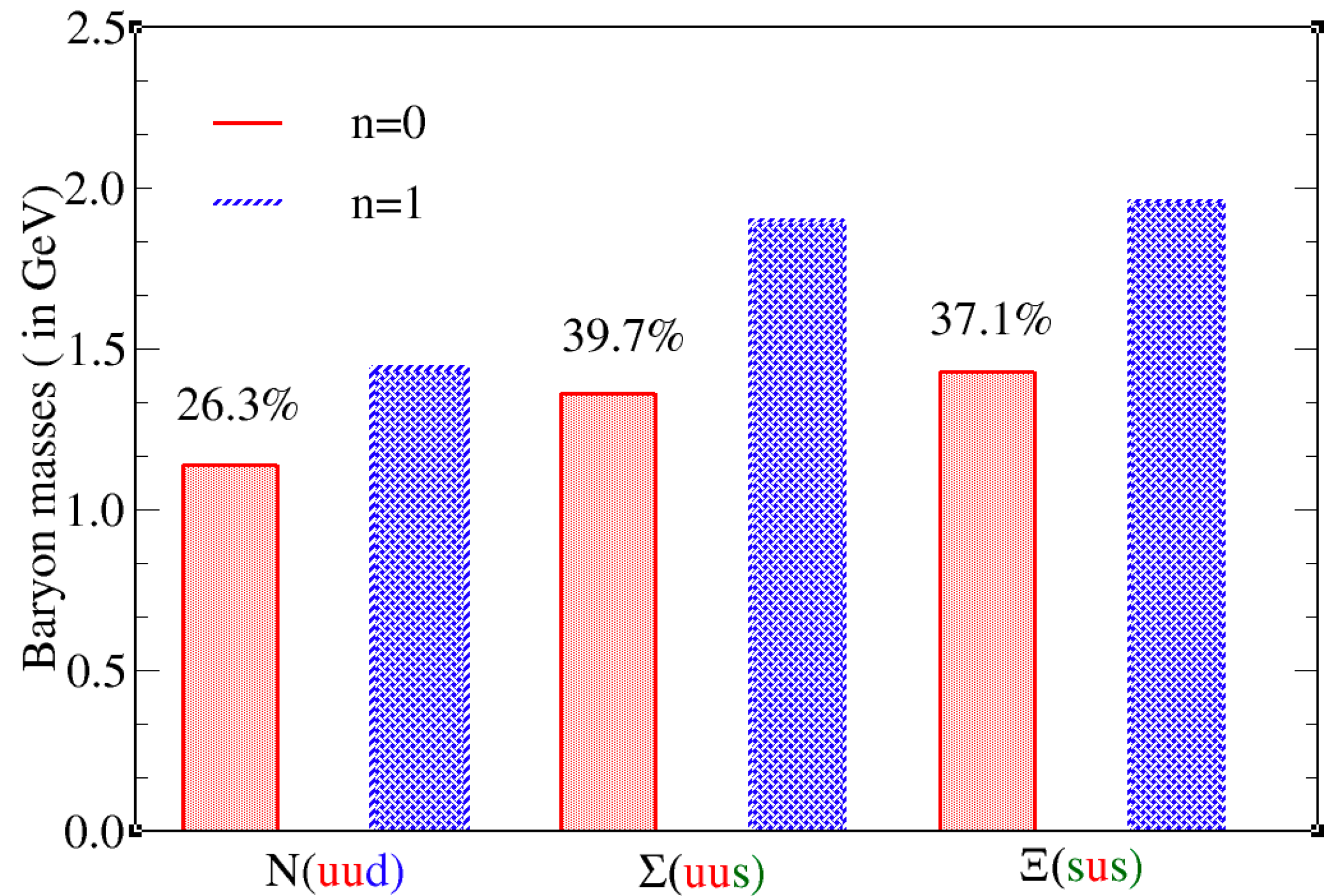}}
       \vspace{-0cm}
\end{figure}

\begin{figure*}[htpb]
\caption{\label{oct-fig} 
Baryons with spin-$1/2$: Comparison between the masses of the ground states and their first radial excitations. Within each pair of bars, the percentage difference between the masses for states $n=0$ and $n=1$ is indicated.}
       \vspace{-0cm}
\centerline{\includegraphics[scale=0.3,angle=0]{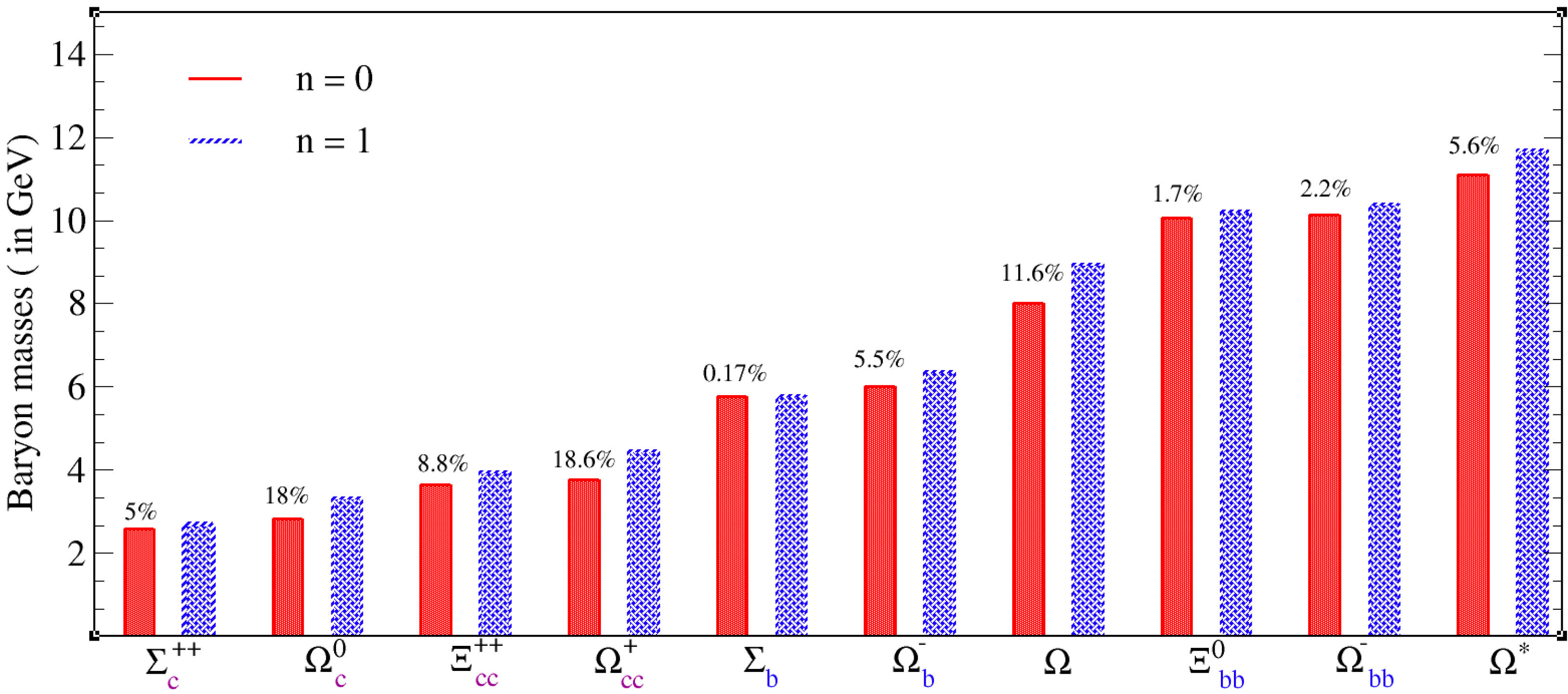}}
       \vspace{-0cm}
\end{figure*}

Experimental and calculated masses for ground and excited light  heavy-light and heavy baryons with spin $1/2$-baryons  are listed in 
Table~\ref{table-baryons-half}, where we have included the resulting masses when varying the parameter $g_B$ in the range $g_B\pm 0.5$
in \eqn{gbs}. 
It is evident from the results shown that the mass decreases when the value of $g_B$
  increases and increases when $g_B$
  decreases.\\
There are no experimental results for the radial excitations of baryons containing one, two, or three heavy quarks. We present our results alongside those obtained in Ref.~\cite{Qin:2019hgk}. \\\
\begin{table}[htbp]
\extrarowheight = -0.5ex
    \renewcommand{\arraystretch}{1.75}
\caption{\label{table-baryons-half} Spin-1/2 baryons.
Masses from CI in the ground state are taken from \cite{Gutierrez-Guerrero:2019uwa,Gutierrez-Guerrero:2021rsx}. Our results for the first radial excitation of the baryons are presented in the sixth column using CI. Additionally, we have included results that explore the variation of the parameter 
$g_B\pm 0.5$.
The results abbreviated by QRS are taken from Ref.~\cite{Qin:2019hgk} and the experimentally results from Ref.~\cite{Workman:2022ynf}.} 
\vspace{0 cm}
\begin{tabular}{@{\extracolsep{0.3 cm}}ccc|ccc}
\hline
\hline
&\multicolumn{2} {c}{$n=0$} & \multicolumn{2} {c}{$n=1$}\\
\hline
Baryon & Exp & CI& Exp & QRS & CI \\
\rule{0ex}{3.5ex}
$N(\tu\tu\td)$ & 0.94&1.14 & 1.44 & 1.44 &$1.44^{-0.11}_{+0.11}$\\
\rule{0ex}{3.5ex}
$\Sigma(\tu\tu\ts)$ &1.19&1.36&1.66 & 1.58 & $1.90^{-0.06}_{+0.08}$\\
\rule{0ex}{3.5ex}
$\Xi(\ts\tu\ts)$ &1.31&1.43&--&1.72 & $1.96^{-0.07}_{+0.09}$\\
\rule{0ex}{3.5ex}
$\Sigma_{\tc}^{++}(\tu\tu\tc)$ &2.45&2.58&--& 2.65 & $2.71^{-0.06}_{+0.07}$
\\ \rule{0ex}{3.5ex}
\rule{0ex}{3.5ex}
$\Omega_{\tc}^0(\ts\ts\tc)$ &2.69&2.82&--& 2.93 & $3.33^{-0.05}_{+0.05}$\\ 
\rule{0ex}{3.5ex}
$\Xi_{cc}^{++} (\tu\tc\tc)$ &3.62&3.64&--&3.86 &$3.96^{-0.14}_{+0.22}$ \\
\rule{0ex}{3.5ex}
$\Omega_{\tc\tc}^+(\ts\tc\tc)$ &--&3.76&--&4.00& $4.46^{-0.09}_{+0.15}$\\
\rule{0ex}{3.5ex}
$\Sigma_b^+(\tu\tu\tb)$ &5.81&5.78&--&5.91 & $5.79^{-0.18}_{+0.15}$\\
\rule{0ex}{3.5ex}
$\Omega_b^{-}(\ts\ts\tb)$ &6.04&6.01&--&6.20 & $6.36$$^{-0.17}_{+0.22}$\\
\rule{0ex}{3.5ex}
$\Omega(\tc\tc\tb)$ &--&8.01&--&8.33 & $8.94^{-0.04}_{+0.04}$\\
\rule{0ex}{3.5ex}
$\Xi_{bb}^0(\tu\tb\tb)$ &--&10.06&--&10.39 & $10.24$$^{-0.03}_{+0.06}$\\ 
\rule{0ex}{3.5ex}
$\Omega_{bb}^-(\ts\tb\tb)$ &--&10.14&--&10.53 & $10.37$$^{-0.12}_{+0.15}$\\ 
\rule{0ex}{3.5ex}
$\Omega(\tc\tb\tb)$ &--&11.09&--&11.60 & $11.71^{-0.04}_{+0.04}$\\
\hline
\hline
\end{tabular}
\end{table}
In Figs.~\ref{oct-lig} and \ref{oct-fig}, we pictorially depict the differences between the ground states and their first radial excitations, emphasizing their percentage variances. It is straightforward to observe that masses of excited states are always more significant than those of the ground states. This is attributed to the zero we have introduced in the kernel and to the fact that  masses of diquarks composing the excited states are more significant than those of the ground states, as evidenced in Table~\ref{fbod}.
In the heavy sector, the maximum difference between the ground and excited states is for $\Omega_{\tc\tc}^+(\ts\tc\tc)$ with 18.6\%, and the minimum is for $\Sigma_{\tb^+}(\tu\tu\tb)$ with 0.17\%.\\
One of our most significant results is the mass obtained for the $\Omega_{\tb}^{-}(\ts\ts\tb)$, as it is consistent with a state recently detected in LHCb as a candidate for the first radial excitation of the $\Omega_{\tb}^{-}(\ts\ts\tb)$  \cite{LHCb:2020tqd}. The detected mass is $6.35$~GeV, and the one we obtain is $6.36$~GeV, yielding a difference of 0.15\% between the two states.\\
Similarly, the candidate for the first radial excitation of $\Omega_{\tc}^0(\ts\ts\tc)$ 
 detected by LHCb shows a mass of $3.18$~GeV \cite{Karliner:2023okv,LHCb:2023sxp,Pan:2023hwt}. Our calculations predict a mass of $3.33$ GeV, resulting in a difference of 4.7\% between the two.\\
In Table~\ref{doublemedio}, we display our results for doubly heavy baryons compared with other models, utilizing the QCD sum rules formalism (SM)\cite{ShekariTousi:2024mso}, Relativistic quark Model (RQM)\cite{Ebert:2002ig},  in a Salpeter model with AdS/QCD inspired potential (SMP) \cite{Giannuzzi:2009gh}.
\begin{table}[htbp]
 \caption{\label{doublemedio} The masses of firts radial excitation of the doubly heavy baryons in GeV and comparison of the results with other predictions.}
    \begin{tabular}{@{\extracolsep{0.3 cm}}c|cccc}
         & CI & SM \cite{ShekariTousi:2024mso} & RQM \cite{Ebert:2002ig}& SMP \cite{Giannuzzi:2009gh} \\
          \hline
         \rule{0ex}{3.5ex}
$\Xi_{cc}^{++} (\tu\tc\tc)$ & 3.96 & 4.0 & 3.91 & 4.18 \\
  \rule{0ex}{3.5ex}
 $\Omega_{\tc\tc}^+(\ts\tc\tc)$ & 4.46 & 4.07 & 4.08 & 4.27 \\ 
  \rule{0ex}{3.5ex}
 $\Xi_{bb}^0(\tu\tb\tb)$ & 10.24 & 10.33 & 10.44 & 10.75 \\
  \rule{0ex}{3.5ex}
 $\Omega_{bb}^-(\ts\tb\tb)$ & 10.37& 10.45 & 10.61 & 10.83\\
  \hline
    \end{tabular}
\end{table}
Our results for triply heavy baryons with spin-$1/2$ are compared in Table~\ref{triplymedio} from SM \cite{Alomayrah:2020qyw}, constituent quark model (QM) \cite{Yang:2019lsg}, the renormalization group procedure for effective particles (EP) \cite{Serafin:2018aih},  and RQM \cite{Faustov:2021qqf,Ebert:2007nw}. Clearly, our results for both states are perfectly consistent with those reported using other models.

\begin{table}[htbp]
 \caption{\label{triplymedio} The masses of first radial excitation of the triply heavy baryons in GeV and comparison of the results with other predictions.}
    \begin{tabular}{@{\extracolsep{0.1 cm}}c|ccccc}
         & CI & SM \cite{Alomayrah:2020qyw} & QM \cite{Yang:2019lsg} & EP \cite{Serafin:2018aih} & RQM \cite{Faustov:2021qqf,Ebert:2007nw}\\
          \hline
         \rule{0ex}{3.5ex}
$\Omega(\tc\tc\tb)$ & 8.94 & 8.65 & 8.46 & 8.6 & 8.40\\
  \rule{0ex}{3.5ex}
$\Omega(\tc\tb\tb)$ & 11.71 & 11.77 & 11.61& 11.59 & 11.69\\
  \hline
    \end{tabular}
\end{table}
The FAs are listed in Table~\ref{FA-oct}, highlighting the dominant diquark in calculating the baryonic mass. At this juncture, we also compare them with the amplitudes obtained for the ground state. From these findings, it becomes apparent that six of the baryons comprised of heavy quarks exhibit altered behavior, wherein the predominant diquark contributing to mass differs between the ground state and the radial excitation. Conversely, only four maintain consistency in behavior.\\
\begin{table*}[htbp]
\caption{\label{FA-oct} Baryon-$1/2$ FA. Notably, the amplitudes give the dominant diquark in the mass of the baryon. In the last column, we indicate this diquark in each case. The baryons that change their dominant diquark upon transitioning from the ground state to the first radial excitation are $N(\tu\tu\td)$, $\Sigma_{\tc}^{++}(\tu\tu\tc)$, $\Xi(\ts\tu\ts)$, $\Omega_{\tc}^0(\ts\ts\tc)$, $\Sigma(\tu\tu\ts)$,  $\Omega(\tc\tc\tb)$, $\Xi_{bb}^0(\tu\tb\tb)$, $\Omega_{\tb\tb}^-(\ts\tb\tb)$ and $\Omega(\tc\tb\tb)$; however, the $\Xi_{cc}^{++} (\tu\tc\tc)$, $\Omega_{\tc\tc}^+(\ts\tc\tc)$,  $\Sigma_{\tb^+}(\tu\tu\tb)$ and  $\Omega_{\tb^{-}}(\ts\ts\tb)$ baryons maintain the same behavior as the ground state
.} 
\vspace{0 cm}
\begin{tabular}{@{\extracolsep{0.4 cm}}ccccccc|c}
\hline
\hline
& & $s$ & $a_1^+$ & $a_1^0$ & $a_2^+$ & $a_2^0$ & dom \\  
 \rule{0ex}{3.0ex}
\multirow{2}{2.0cm}{$N(\tu\tu\td)$ } &$ n=0$ &
-0.02 &
0.52 &
-0.37 &
-0.63 &
0.44 &
$\{\tu\tu\}\td$ 
\\ \rule{0ex}{3.0ex}
& $n=1$ & -0.76&
-0.11 &
0.14 &
0.50 &
-0.36 &
$[\tu\td]\tu$ 
\\
\hline
\rule{0ex}{3.0ex} 
\multirow{2}{2.0cm}{$\Sigma(\tu\tu\ts)$} & $n=0$ 
& 0&
-0.04 &
0.02 &
0.83 &
-0.55 &
$\{\tu\tu\}\ts$ 
\\ \rule{0ex}{3.0ex}
& $n=1$ & -0.85&
-0.20 &
0.18 &
0.37 &
-0.22 &
$[\tu\ts]\tu$ 
\\
\hline
\rule{0ex}{3.0ex} 
\multirow{2}{2.0cm}{$\Xi(\ts\tu\ts)$} & $n=0$ 
& 0&
0.01 &
-0.99 &
-0.02 &
0.06 &
$\{\tu\ts\}\ts$ 
\\ \rule{0ex}{3.0ex}
& $n=1$ & 0.79&
0.21 &
-0.18 &
-0.47 &
0.23 &
$[\tu\ts]\ts$ 
\\
\hline
\rule{0ex}{3.0ex} 
\multirow{2}{2.0cm}{$\Sigma_{\tc}^{++}(\tu\tu\tc)$ } &
$n=0$ & -0.49&
0.26 &
-0.09 &
-0.82 &
0.01 &
$\{\tu\tu\}\tc$ 
\\ \rule{0ex}{3.0ex}
& $n=1$ & 0.94&
0.27 &
-0.09 &
0.12 &
0.03 &
$[\tu\tc]\tu$
\\
\hline
\rule{0ex}{3.0ex} 
\multirow{2}{2.0cm}{$\Omega_{\tc}^0(\ts\ts\tc)$}& $n=0$  & 0.57 &
-0.29 &
0.10 &
0.76 &
-0.02 &
$\{\ts\ts\}\tc$ 
\\
\rule{0ex}{3.0ex}
& $n=1$  & 0.92 &
0.29 &
-0.15 &
0.17 &
0.07 &
$[\ts\tc]\ts$
\\
\hline
\rule{0ex}{3.0ex} 
\rule{0ex}{3.0ex} 
 \multirow{2}{2.0cm}{$\Xi_{cc}^{++} (\tu\tc\tc)$ } & $n=0$ & -0.88 & 0.05 &
-0.36&
0.11 &
0.30 & $[\tu\tc]\tc$ \\
\rule{0ex}{3.0ex} 
& $n=1$  & -0.93 &
-0.04 &
0.22 &
0.05 &
-0.25 & $[\tu\tc]\tc$
\\
\hline
\rule{0ex}{3.0ex} 
\multirow{2}{2.0cm}{$\Omega_{\tc\tc}^+(\ts\tc\tc)$  } & $n=0$ & -0.88 &
0.07 &
-0.35 &
0.11 &
0.30 & $[\ts\tc]\tc$ 
\\ \rule{0ex}{3.0ex}
& $n=1$  & 0.90 &
0.09 &
-0.31 &
0.11 &
0.25 & $[\ts\tc]\tc$ 
\\
\hline
\rule{0ex}{3.0ex} 
  \multirow{2}{2.0cm}{  $\Sigma_b^+(\tu\tu\tb)$ } & $n=0$& 0.5 &
-0.20&
0.05 &
0.83 &
0.04 &
$\{\tu\tu\}\tb$  
\\ \rule{0ex}{3.0ex}
& $n=1$& 0.65 &
0.26&
0.01 &
0.71 &
-0.02 &
$\{\tu\tu\}\tb$ 
\\
\hline
 \rule{0ex}{3.0ex} 

 \rule{0ex}{3.0ex} 
 \multirow{2}{2.0cm}{  $\Omega_b^{-}(\ts\ts\tb)$} & $n=0$ & 0.12 &
-0.10 &
0.04 &
-0.98 &
-0.07 &
$\{\ts\ts\}\tb$ 
\\ \rule{0ex}{3.0ex}
& $n=1$ &  0.68 &
 0.22 &
 0.01&
 0.69&
-0.03 &
$\{\ts\ts\}\tb$ 
\\
\hline
 \rule{0ex}{3.0ex} 
  \multirow{2}{2.0cm}{ $\Omega(\tc\tc\tb)$ }& $n=0$& -0.82 &
0.21&
-0.008&
-0.53&
-0.003&
$[\tc\tb]\tc$  
\\ \rule{0ex}{3.0ex}
& $n=1$& 0.61 &
0.38&
0.19&
0.65&
0.08&
$\{\tc\tc\}\tb$ 
\\
\hline
\rule{0ex}{3.0ex} 
\multirow{2}{2.0cm}{$\Xi_{bb}^0(\tu\tb\tb)$ }& $n=0$ & -0.11 &
0.07 &
-0.04 &
0.99 &
0.06 &
$\{\tb\tb\}\tu$ 
\\ \rule{0ex}{3.0ex}
& $n=1$ & 0.98 &
 0.008 &
-0.12 &
0.005 &
0.13 &
$[\tu\tb]\tb$ 
\\
\hline
\rule{0ex}{3.0ex} 
\multirow{2}{2.0cm}{ $\Omega_{bb}^-(\ts\tb\tb)$} & $n=0$ & 0.12 &
-0.10 &
0.04 &
-0.98 &
-0.07 &
$\{\tb\tb\}\ts$ 
\\ \rule{0ex}{3.0ex}
& $n=1$  &
-0.96 &
-0.01 &
0.16 &
-0.07 &
-0.16 &
$[\ts\tb]\tb$ 
\\
\hline
%
 \rule{0ex}{3.0ex} 
  \multirow{2}{2.0cm}{ $\Omega(\tc\tb\tb)$ }  & $n=0$& -0.77&
0.05&
-0.30 &
0.49&
0.28&
$[\tc\tb]\tb$  
\\ \rule{0ex}{3.0ex}
& $n=1$& 
0.59&
0.07 &
-0.17&
0.72&
0.28 &
$\{\tc\tb\}\tb$ 
\\
\hline
\hline
\hline
\end{tabular}
\end{table*}
The masses of radial excitations, as well as the masses in the ground state of the baryons, must adhere to the spacing rule.
Light baryons satisfy the Gell-Mann–Okubo mass formula \cite{GellMann:1962xb,Okubo:1961jc}:
\bea
\label{gomf}
\frac{1}{2}(m_{N(\tu\tu\td)}+ m_{\Xi(\ts\tu\ts)})=\frac{1}{4}(3m_{\Lambda(\tu\td\ts)}+m_{\Sigma(\tu\tu\ts)})\,.
\eea
that predicts a mass of 1.63~GeV for $m_{\Lambda(\tu\td\ts)}$, which is in remarkably good agreement with the experimental result of 1.60~GeV \cite{ParticleDataGroup:2010dbb}.
For baryons with one heavy quark, the spacing rule satisfies~\cite{Ebert:2005xj,GellMann:1962xb,Okubo:1961jc},
\bea
&&m_{\Sigma_Q}+m_{\Omega_Q}=2m_{\Xi_Q} \;, \;\;\;\;\;\;\;\;\;Q=\tc,\;\tb\;. 
\eea
We test this rule for the following baryons~:
\begin{eqnarray}
\label{gm-l1} m_{\Sigma_{\tc}^{++}(\tu\tu\tc)}+m_{\Omega_{\tc}^0(\ts\ts\tc)} &=& 2m_{\Xi_{\tc}^+{(\tu\ts\tc)}},\\
\label{gm-l2} m_{\Sigma_{\tb}^+{(\tu\tu\tb)}}+m_{\Omega_{\tb}^-{(\ts\ts\tb)}} &=& 2m_{\Xi_{\tb}^0{(\tu\ts\tb)}} \;.
\end{eqnarray}
Using Eqs.~(\ref{gm-l1})-(\ref{gm-l2}) and the results in Table \ref{table-baryons-half}, we find $m_{\Xi_{\tc}^+{(\tu\ts\tc)}}=3.02$ $\GeV$ and $m_{\Xi_{\tb}^0{(\tu\ts\tb)}}= 6.125$ $\GeV$. 
Applying the identical calculation to the predictions derived within the QRS framework yields $m_{\Xi_{\tc}^{+}(\tu\ts\tc)}=2.79$ GeV and  $m_{\Xi_{\tb}^{0}{(\tu\ts\tb)}}=6.055$ GeV. For the proposed state of the first radial excitation of $\Xi_{\tb}^0(\tu\ts\tb)$, the mass found by LHCb in Refs.~\cite{LHCb:2018vuc,Aliev:2018lcs} is $6.22$ GeV, remarkably close to the result derived from our model.\\
\subsection{Fisrt Radial Excitation: Baryons with spin-3/2}
\begin{figure}[htpb]
\caption{\label{decuplet-lig}Comparison between the masses of the ground states  and their first radial excitations for light baryons.}
       \vspace{0cm}
       \centerline{\includegraphics[scale=0.23,angle=0]{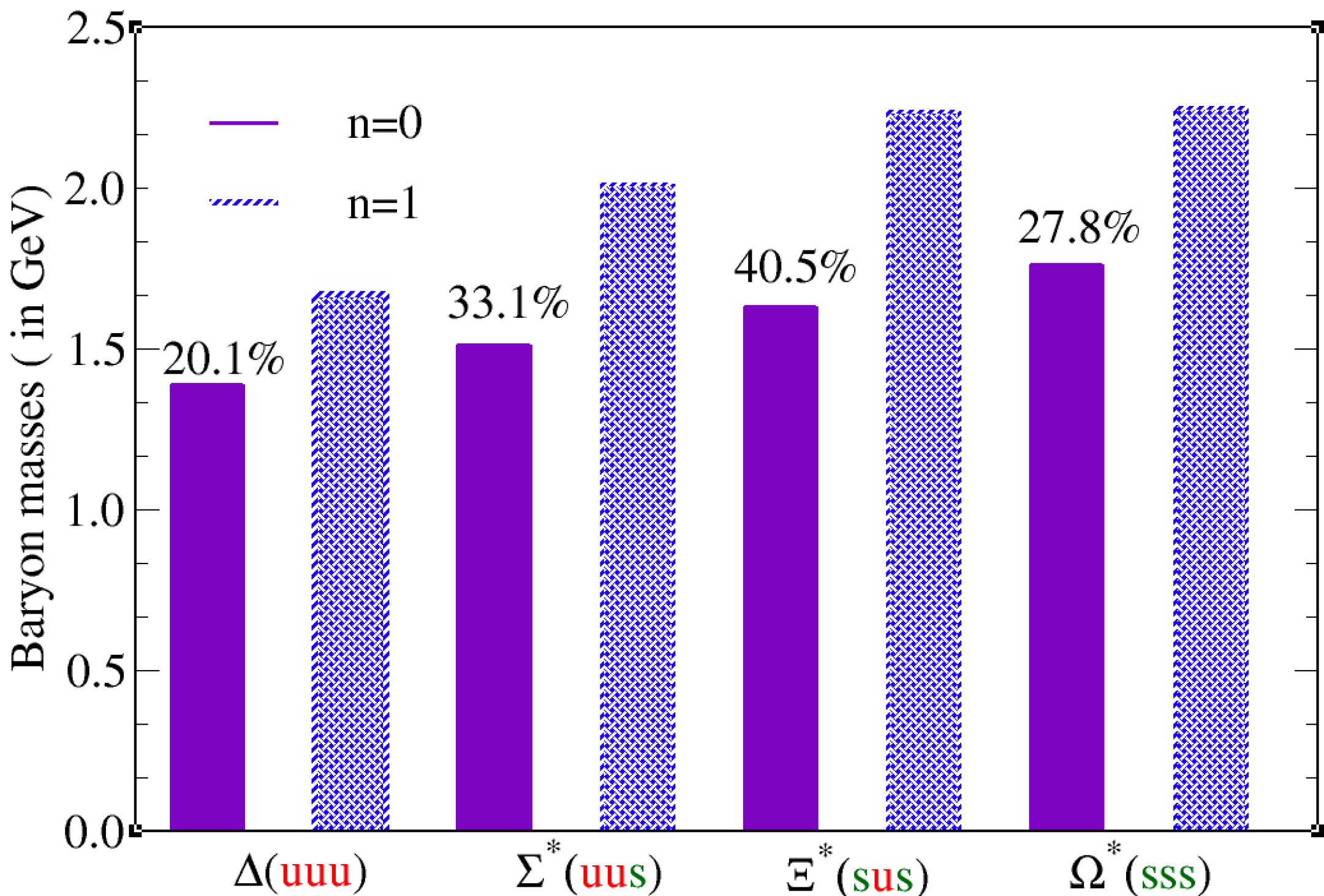}}
       \vspace{0cm}
\end{figure}

\begin{figure*}[htpb]
\caption{\label{decuplet-fig-2}Baryons with spin-3/2:  Comparison between the masses of their ground states  and their first radial excitations. Within each pair of bars, the percentage difference between the masses for states $n=0$ and $n=1$ is indicated.}
       \vspace{-0cm}
       \centerline{\includegraphics[scale=0.3,angle=0]{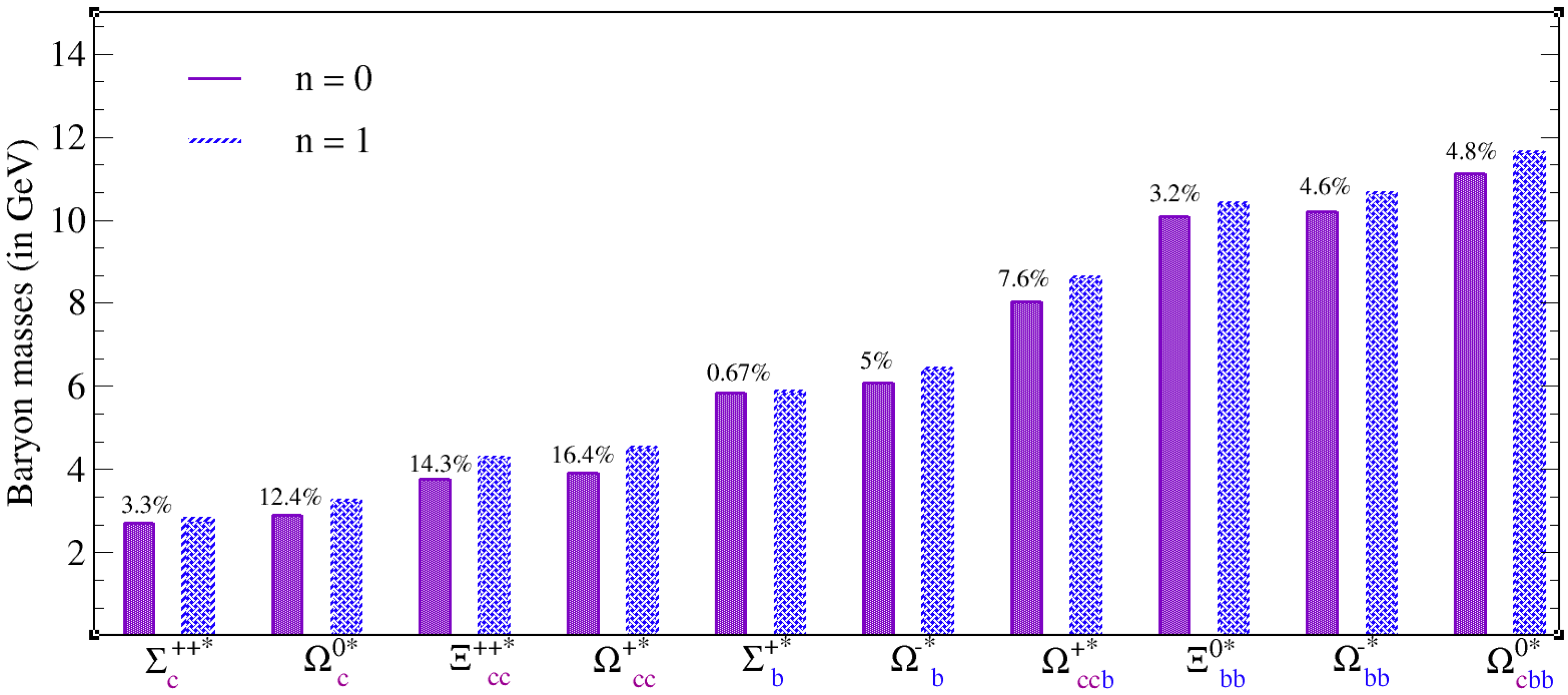}}
       \vspace{-0cm}
\end{figure*}

\begin{figure}[htpb]
\caption{\label{decuplet-omegac}Comparison between the masses of the ground states  and their first radial excitations for $\Omega^{++*}_{\tc\tc\tc}$.}
       \vspace{0cm}       \centerline{\includegraphics[scale=0.23,angle=0]{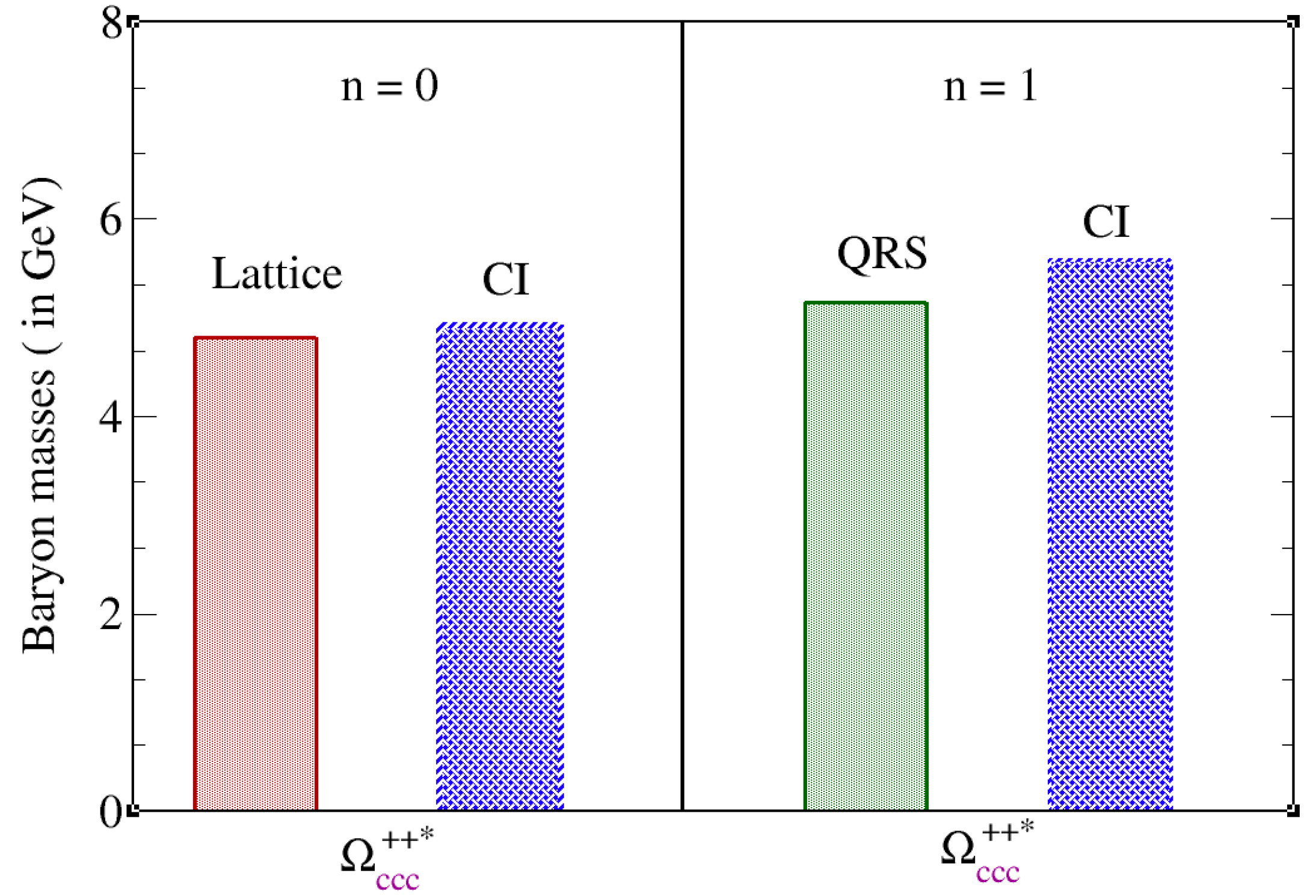}}
       \vspace{0cm}
\end{figure}

\begin{figure}[htpb]
\caption{\label{decuplet-omegab}Comparison between the masses of the ground states  and their first radial excitations for $\Omega^{-*}_{\tb\tb\tb}$.}
       \vspace{0cm}  \centerline{\includegraphics[scale=0.23,angle=0]{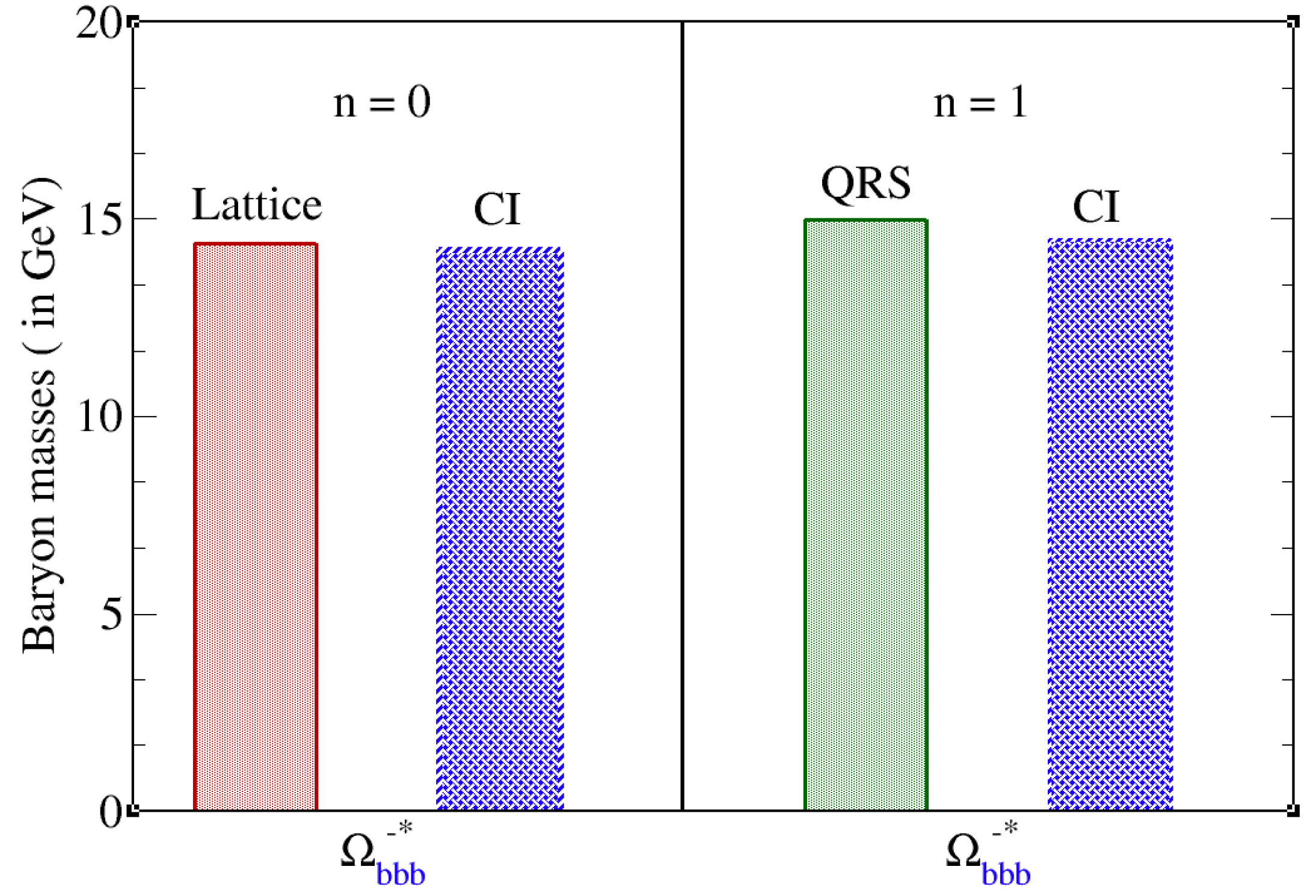}}
       \vspace{0cm}
\end{figure}

Now we focus our attention on the heaviest baryons, which include baryons containing three identical heavy quarks in their configuration.\\
We show our results for the first radial excitation of these states in Table~\ref{table-baryons-dec},  
 where, as in the case of spin-1/2 baryons, we have analyzed our results by varying the parameter $g_B$. We found that, as in the previous case, the mass increases when 
$g_B$ decreases and decreases when $g_B$
  increases. We also compared our results with known experimental data \cite{Workman:2022ynf}, the QRS model \cite{Qin:2019hgk}, and the ground states calculate by Lattice in Refs.~\cite{Brown:2014ena,Mathur:2018epb} and CI in Refs.~\cite{Gutierrez-Guerrero:2019uwa,Gutierrez-Guerrero:2021rsx}.
\begin{table}[htbp]
\extrarowheight = -0.5ex
    \renewcommand{\arraystretch}{1.75}
\caption{\label{table-baryons-dec} Masses of baryons with spin-3/2. 
Masses from CI in the ground state are taken from Refs.~\cite{Gutierrez-Guerrero:2019uwa,Gutierrez-Guerrero:2021rsx}.
Our results for the first radial excitation with CI are shown in the last column, where we have included the results of varying the parameter $g_B\pm 0.5$.
Results denoted by QRS have been taken from Ref.~\cite{Qin:2019hgk}, lattice (Lat.) values are taken from \cite{Brown:2014ena,Mathur:2018epb}, and the experimentally results from Ref.~\cite{Workman:2022ynf}.} 
\vspace{0 cm}
\begin{tabular}{@{\extracolsep{0.1 cm}}cccc|ccc}
\hline
\hline
& \multicolumn{3}{c}{$n=0$} &  \multicolumn{3}{c}{$n=1$}\\
\hline
Baryon & Lat. & CI  & Exp.  & Exp. & QRS  & CI\\
$\Delta(\tu\tu\tu)$ &1.23&1.39&1.65 & 1.60 & 1.46 & $1.67^{-0.07}_{+0.03}$\\
\rule{0ex}{3.5ex}
$\Sigma^*(\tu\tu\ts)$ &1.39&1.51&1.67&1.73& 1.627 & $2.01^{-0.02}_{+0.02}$\\
\rule{0ex}{3.5ex}
$\Xi^*(\ts\tu\ts)$ &1.53&1.63&1.82&--&1.793 & $2.24^{-0.04}_{+0.05}$\\
\rule{0ex}{3.5ex}
$\Omega(\ts\ts\ts)$ &1.67&1.76&--&--& 1.96 & $2.25^{-0.04}_{+0.03}$\\
\hline
\hline
 $\Sigma_{\tc}^{++*}{(\tu\tu\tc)}$ &2.52&2.71&--&--&2.80 & $2.80^{-0.02}_{+0.04}$\\
  \rule{0ex}{3.5ex}
 $\Omega_{\tc}^{0*}{(\ts\ts\tc)}$  &2.76&2.90&--&--&3.02 &$3.26^{-0.02}_{+0.05}$ \\
 \rule{0ex}{3.5ex}
 $\Xi_{\tc\tc}^{++*}{(\tu\tc\tc)}$  &3.57&3.76&--&--&3.97 & $4.30^{-0.02}_{+0.03}$\\
 \rule{0ex}{3.5ex}
 $\Omega_{\tc\tc}^{+*}{(\ts\tc\tc)}$ &3.71&3.90&--&--&4.08 & $4.54^{-0.01}_{+0.04}$\\
  \rule{0ex}{3.5ex}
 $\Sigma_{\tb}^{+*}{(\tu\tu\tb)}$ &5.75&5.85&--&--&6.07 & $5.89$$^{-0.04}_{+0.03}$\\
  \rule{0ex}{3.5ex}
   $\Omega_{\tb}^{-*}{(\ts\ts\tb)}$&5.99&6.09&--&--&6.30 & $6.41$$^{-0.03}_{+0.01}$\\
  \rule{0ex}{3.5ex}
 $\Xi_{bb}^{0*}{(\tu\tb\tb)}$ &10.04&10.09&--&--&10.52& $10.42^{-0.01}_{+0.04}$\\
  \rule{0ex}{3.5ex}
 $\Omega_{\tb\tb}^{-*}{(\ts\tb\tb)}$  &10.18&10.20&--&--& 10.64 &$10.67^{-0.02}_{+0.01}$\\
 \hline
\hline
 $\Omega_{\tc\tc\tc}^{++*}$ &4.80&4.93&-- &--&5.15 & $5.59^{-0.03}_{+0.01}$\\
   \rule{0ex}{3.5ex}
 $\Omega_{\tc\tc\tb}^{+*}$  &8.01&8.03&--&--&8.42 &$8.64^{-0.02}_{+0.01}$\\
  \rule{0ex}{3.5ex}
 $\Omega_{\tc\tb\tb}^{0*}$  &11.20&11.12&--&--&11.70 & $11.66^{-0.01}_{+0.02}$\\ 
  \rule{0ex}{3.5ex}
 $\Omega_{\tb\tb\tb}^{-*}$  &14.37&14.23&--&--&14.98 &$14.66^{-0.01}_{+0.01}$\\
 \hline 
  \hline
\end{tabular}
\end{table}
In Figs.~\ref{decuplet-lig} and~\ref{decuplet-fig-2}, we have plotted the masses reported in Table~\ref{table-baryons-dec} to represent these values visually, similar to the case of baryons with spin-$1/2$, along with their percentage differences.
It is readily apparent that all radial excitations surpass their ground states. Furthermore, in the
heavy sector  the largest discrepancy is observed for  $\Omega_{\tc\tc}^{+*}{(\ts\tc\tc)}$ at 16.4\%, while the smallest is seen for $\Sigma_{\tb}^{+*}{(\tu\tu\tb)}$ at 0.67\%.
Of particular significance is the outcome for $\Omega_{\tc}^{0*}$, LHCb proposes a mass of 3.3~GeV for this state in Refs.~\cite{Karliner:2023okv,LHCb:2023sxp,Pan:2023hwt}, while our CI model yields 3.26~GeV presenting a small difference of approximately 1.21\%.\\
In Table~\ref{pesados-decu-2}, we present the heaviest baryons with spin-3/2, comprised of two and three $\tb$ and $\tc$ quarks. Additionally, we also display results for these baryons obtained using alternative models,
QRS \cite{Qin:2019hgk}, RQM \cite{Faustov:2021qqf,Ebert:2007nw} and Regge phenomenology (RF) \cite{Oudichhya:2023pkg,Shah:2017jkr}.
\begin{table}[htbp]
\extrarowheight = -0.5ex
    \renewcommand{\arraystretch}{1.75}
\caption{\label{pesados-decu-2} Masses of triple heavy baryons with spin-$3/2$, compared to the masses predicted by other models.} 
\vspace{0 cm}
\begin{tabular}{@{\extracolsep{0.1 cm}}ccccc}
\hline
\hline
Baryon & CI & QRS \cite{Qin:2019hgk} & RQM \cite{Faustov:2021qqf,Ebert:2007nw} & RF \cite{Oudichhya:2023pkg,Shah:2017jkr}\\
 $\Omega_{\tc\tc\tb}^{+*}$  & 8.64 &8.42 & 8.41 & 8.63\\
  \rule{0ex}{3.5ex}
 $\Omega_{\tc\tb\tb}^{0*}$  &11.66 & 11.70 & 11.70 &11.79  \\ 
  \rule{0ex}{3.5ex}
  $\Omega_{\tc\tc\tc}^{++*}$ & 5.59 & 5.15 & 5.54 &5.3\\
   \rule{0ex}{3.5ex}
 $\Omega_{\tb\tb\tb}^{-*}$ &14.66 & 14.98 & 15.12&15.16\\
 \end{tabular}
 \end{table}
To conduct a visual analysis of these baryons within Figs.~\ref{decuplet-omegac} and \ref{decuplet-omegab}, we delineate the triple baryons composed of $\tc\tc\tc$ and $\tb\tb\tb$. The left panel illustrates the baryon ground states, while the right panel depicts their first radial excitation. This graphical representation facilitates a direct comparison with alternative approaches.\\
In Tables~\ref{table-baryons-dec} and \ref{pesados-decu-2}, we emphasize that the results obtained with CI in the ground state and the first excited state are consistent with other models, such as lattice and QRS.\\
We define a constituent-quark passive-mass in analogy with the ground state of baryons~\cite{Qin:2018dqp,Gutierrez-Guerrero:2019uwa}, via 
\bea
\label{EqMfP}
M_{f}^{P} &=& \frac{1}{3} m_{\Omega_{fff}}\,.
\eea
In Table \ref{Pm}, we compare the computed values (in GeV) from this relation
with the input parameters we used for CI in the ground and first radial excitation. This mass increases when we transition from $n=0$ to $n=1$, just as expected.\\
\begin{table}[htbp]
\caption{\label{Pm}Constituent-quark passive-mass derived from CI, Lattice, and the QRS model, corresponding to states $n=0$ and $n=1$.}
\begin{tabular}{@{\extracolsep{0.3cm}}ccc|cc}
\hline
&&& $\tc$ & $\tb$ \\
\hline
  \rule{0ex}{3.5ex}
& & $M_{f}$  & $1.53$ & $4.68$ \\
\hline
  \rule{0ex}{3.5ex}
\multirow{2}{0.5cm}{$n=0$}
& \text{CI} & $M_{f}^{P}$ & 1.64 & 4.74 \\
 \rule{0ex}{3.5ex}
& \text{Lat.} & $M_f^P$    &1.6 & 4.79 \\
\hline
\rule{0ex}{3.5ex}
\multirow{2}{0.5cm}{$n=1$}
& \text{CI} & $M_f^P$ & 1.86  & 4.88 \\
 \rule{0ex}{3.5ex}
& \text{QRS} &  $M_f^P$ & 1.71 & 4.99 \\
\hline
\hline
\end{tabular}
\end{table}

The masses of baryons with spin-$3/2$ with a single heavy quark obey an equal-spacing rule~\cite{Ebert:2005xj,GellMann:1962xb,Okubo:1961jc}
\bea \label{gmo-h}
m_{\Sigma_Q}+m_{\Omega_Q}=2m_{\Xi_Q} \,.\;\;\;\;\;\;\;\;\;Q=\tc,\;\tb
\eea
This relation yields a mass of $m_{\Xi_{\tu\ts\tc}}=3.03$ GeV. For the corresponding baryon containing a bottom quark, we obtain $m_{\Xi_{\tu\ts\tb}}=6.19$ GeV.\\
We now turn our attention to the spacing rules which combine baryons with different 
spins~\cite{Yu:2018com}~:
\bea \label{gmo-10}
\hspace{-1cm}&&m_{\Xi_{\tc\tc}^{++*}{(\tu\tc\tc)}}\hspace{-0.1cm}-\hspace{-0.09cm}m_{\Xi_{\tc\tc}^{++}(\tu\tc\tc)}\hspace{-0.1cm}-\hspace{-0.09cm}m_{\Sigma_{\tc}^{++*}{(\tu\tu\tc)}}\hspace{-0.1cm}+\hspace{-0.1cm}m_{\Sigma_{\tc}^{++}(\tu\tu\tc)}\hspace{-0.09cm}=0\,,\\
\label{gmo-11}\hspace{-1cm}&&m_{\Omega_{cc}^{+*}{(\ts\tc\tc)}}-m_{\Omega_{\tc\tc}^{+}(\ts\tc\tc)}-m_{\Omega_{\tc}^{0*}{(\ts\ts\tc)}}+m_{\Omega_{\tc}^{0}(\ts\ts\tc)}=0 \,,\\
\label{gmo-12}\hspace{-1cm}&&m_{\Xi_{\tb\tb}^{0*}{(\tu\tb\tb)}}-m_{\Xi_{\tb\tb}^{0}(\tu\tb\tb)}-m_{\Sigma_{\tb}^{+*}{(\tu\tu\tb)}}+m_{\Sigma_{\tb}^{+}(\tu\tu\tb)}=0 \,.
\eea
Using the results for the masses from Tables~\ref{table-baryons-half} and \ref{table-baryons-dec} for CI, we obtain, 0.25 GeV, 0.15 GeV and 0.08 GeV for Eqs.~(\ref{gmo-10})-(\ref{gmo-12}) respectively. 
For completeness and enhanced analysis of the first radial excitations of baryons composed of heavy quarks, we enumerate the FAs in Table~\ref{table-eigenvector} while also highlighting the dominant diquark in each baryon.\\
\begin{table}[htbp]
\begin{center}
\caption{\label{table-eigenvector}FAs for spin-$3/2$ baryons.
According to our analysis, the dominant diquarks are listed in the last column. The baryons that change their dominant diquark upon transitioning from the ground state to the first radial excitation are 
$\Omega_{\tc\tc}^{+*}{(\ts\tc\tc)}$,
$\Sigma_{\tb}^{+*}{(\tu\tu\tb)}$,
$\Omega_{\tb}^{-*}{(\ts\ts\tb)}$
$\Omega_{\tc\tc\tb}^{+*}$,
; however, the 
$\Sigma^*(\tu\tu\ts)$,
$\Sigma^*(\tu\tu\ts)$,
$\Omega_{\tc}^{0*}{(\ts\ts\tc)}$, 
$\Xi_{\tc\tc}^{++*}{(\tu\tc\tc)}$, 
$\Sigma_{\tc}^{++*}{(\tu\tu\tc)}$,
$\Xi_{\tb\tb}^{0*}{(\tu\tb\tb)}$,
$\Omega_{\tb\tb}^{-*}{(\ts\tb\tb)}$, 
 and $\Omega_{\tc\tb\tb}^{0*}$
baryons maintain the same behavior as the ground state.}
\begin{tabular}{@{\extracolsep{0.3cm}}ccccc}
\hline \hline
&& $d^{\{qq_1\}}$ & $d^{\{qq\}}$ & dom. \\
 \rule{0ex}{3.0ex} 
\rule{0ex}{3.0ex} 
 \multirow{2}{1.5cm}{$\Sigma^*(\tu\tu\ts)$} & $n=0$ &
0.25 &
0.96 &
 $\{\tu\tu\}\ts$ 
\\ \rule{0ex}{3.0ex}
& $n=1$ &
-0.67 &
-0.73 &
 $\{\tu\tu\}\ts$ \\
\hline
%
%
\rule{0ex}{3.0ex} 
\rule{0ex}{3.0ex} 
 \multirow{2}{1.5cm}{$\Xi^*(\ts\tu\ts)$} & $n=0$ &
0.42 &
0.90 &
 $\{\ts\ts\}\tu$ 
\\ \rule{0ex}{3.0ex}
& $n=1$ &
-0.68 &
-0.72 &
 $\{\ts\ts\}\tu$ \\
\hline
%
%
\rule{0ex}{3.0ex} 
\rule{0ex}{3.0ex} 
 \multirow{2}{1.5cm}{$\Sigma_{\tc}^{++*}{(\tu\tu\tc)}$} & $n=0$ &
-0.67 &
-0.74 &
 $\{\tu\tu\}\tc$ 
\\ \rule{0ex}{3.0ex}
& $n=1$ &
-0.68 &
-0.72 &
 $\{\tu\tu\}\tc$ \\
\hline
%
%
\rule{0ex}{3.0ex} 
 \multirow{2}{1.5cm}{$\Omega_{\tc}^{0*}{(\ts\ts\tc)}$} & $n=0$ &-0.60 &
-0.80& $\{\ts\ts\}\tc$
\\ \rule{0ex}{3.0ex}
& $n=1$ &-0.67&
-0.73& $\{\ts\ts\}\tc$ \\
\hline
%
\rule{0ex}{3.0ex} 
 \multirow{2}{1.5cm}{$\Xi_{\tc\tc}^{++*}{(\tu\tc\tc)}$} & $n=0$&
-0.12 &
-0.99 &
 $\{\tc\tc\}\tu$ 
\\ \rule{0ex}{3.0ex}
& $n=1$ &
-0.68 &
-0.73 &
 $\{\tc\tc\}\tu$ \\
\hline
\rule{0ex}{3.0ex} 
 \rule{0ex}{3.0ex} 
 \multirow{2}{1.5cm}{$\Omega_{\tc\tc}^{+*}{(\ts\tc\tc)}$} & $n=0$ &-0.17 &
-0.99 & $\{\tc\tc\}\ts$ 
\\ \rule{0ex}{3.0ex}
& $n=1$ &-0.73 &
-0.67 & $\{\tc\ts\}\tc$ \\
\hline
 \rule{0ex}{3.0ex} 
 \multirow{2}{1.5cm}{$\Sigma_{\tb}^{+*}{(\tu\tu\tb)}$} & $n=0$ &
-0.56 &
-0.83 &
 $\{\tu\tu\}\tb$
\\ \rule{0ex}{3.0ex}
& $n=1$ &
-0.79 &
-0.60 &
 $\{\tu\tb\}\tu$ \\
\hline
%
%
%
\rule{0ex}{3.0ex} 
 \multirow{2}{1.5cm}{$\Omega_{\tb}^{-*}{(\ts\ts\tb)}$} & $n=0$ &
-0.42 &
-0.90 &
 $\{\ts\ts\}\tb$
\\ \rule{0ex}{3.0ex}
& $n=1$ &
 -0.73 &
-0.67 &
 $\{\ts\tb\}\ts$ \\
\hline
%
\rule{0ex}{3.0ex} 
 \multirow{2}{1.5cm}{$\Xi_{\tb\tb}^{0*}{(\tu\tb\tb)}$}& $n=0$ &
0.03 &
0.99 &
 $\{\tb\tb\}\tu$ 
\\ \rule{0ex}{3.0ex}
& $n=1$ &
-0.34 &
-0.9 &
 $\{\tb\tb\}\tu$ \\
 \hline
\rule{0ex}{3.0ex} 
 \multirow{2}{1.5cm}{$\Omega_{\tb\tb}^{-*}{(\ts\tb\tb)}$}& $n=0$ &
-0.02 &
-0.99 &
 $\{\tb\tb\}\ts$
\\ \rule{0ex}{3.0ex}
& $n=1$ &
-0.40 &
-0.91 &
 $\{\tb\tb\}\ts$ \\
\hline
 \rule{0ex}{3.0ex} 
  \multirow{2}{1.5cm}{ $\Omega_{\tc\tc\tb}^{+*}$ }& $n=0$ & -0.35 & -0.93 &$\{\tc\tc\}\tb$
\\ \rule{0ex}{3.0ex}
& $n=1$ & -0.92 & -0.38 &$\{\tc\tb\}\tc$ 
\\
\hline
%
 \rule{0ex}{3.0ex} 
 \multirow{2}{1.5cm}{$\Omega_{\tc\tb\tb}^{0*}$ } & $n=0$& -0.16 &
-0.99 & $\{\tb\tb\}\tc$
\\ \rule{0ex}{3.0ex}
& $n=1$& -0.22 &
-0.97 & $\{\tb\tb\}\tc$ \\
\hline
%
%
\hline
\end{tabular}
\end{center}
\end{table}
Analogously to the case of baryons with spin-1/2, baryons with spin-3/2 change the dominant diquark in four cases when transitioning from the ground state to the excited state.
%
%
\section{Conclusions}
\label{Conclusions}
In this paper, we utilize the CI model to compute the mass spectra of singly, doubly, and triply heavy baryons. The approach includes a quark-diquark approximation to simplify the complex three-body problem into two simpler two-body problems.
In this respect, we employed the excited diquarks previously obtained with CI in Ref. \cite{Paredes-Torres:2024mnz} and integrated them into the Faddeev equation described in Section \ref{Baryons}, \fig{faddevv-Fig1}.
In Tables~\ref{table-baryons-half} and~\ref{table-baryons-dec}, we present the masses predicted by our model using the parameters described throughout our article and in the appendices. 
These predictions are compared in Figs. \ref{oct-fig} and \ref{decuplet-fig-2} with the masses in the ground estate.
Additionally, in Tables~\ref{FA-oct} and \ref{table-eigenvector}, we have included the FAs for all the baryons studied here. \\
The predictions obtained in this work with CI for light baryons with spin $1/2$ satisfy the well-known  Gell-Mann–Okubo mass formula \cite{GellMann:1962xb,Okubo:1961jc} in \eqn{gomf}.
In the case of baryons with a singly heavy quark ($\Sigma_{\tc}^{++},\Omega_{\tc}^0,
\Sigma_{\tb}^+,\Omega_{\tb}^-,\Sigma_{\tc}^{++*},\Omega_{\tc}^{0*},
\Sigma_{\tb}^{+*},\Omega_{\tb}^{-*}$) our results are consistent with those detected by LHCb \cite{LHCb:2018vuc,LHCb:2020tqd,LHCb:2023sxp} and satisfy the mass spacing rule~\cite{Ebert:2005xj,GellMann:1962xb,Okubo:1961jc} in Eqs. (\ref{gm-l1}), (\ref{gm-l2}) and (\ref{gmo-h}).
For baryons with spin-$1/2$ and double heavy quarks, we compare results obtained using other models in Table~\ref{doublemedio}. Results are very close and consistent with existing theoretical predictions.
Our results for baryons with two heavy quarks with different spins also satisfy the relations in the Eqs.~(\ref{gmo-10})-(\ref{gmo-12}). 
Tables \ref{triplymedio} and \ref{pesados-decu-2} compare the results obtained here for triply heavy baryons with those obtained using other approaches. Furthermore, in Figs. \ref{decuplet-omegac} and \ref{decuplet-omegab}, we plot the masses of baryons composed of three quarks $\tb$ and $\tc$, comparing them with other models and their ground states.
In conclusion, we computed the masses of radial excitations for twenty states composed of heavy quarks and provided the corresponding FAs for each state. This lays the groundwork for subsequent calculations of form factors, charge radii, decay constants, and other observables for baryons.
Also, this study will undoubtedly assist future experimental investigations in identifying baryonic states through resonances.
\begin{acknowledgements}
The authors thank Professor Adnan Bashir for the invaluable discussions and suggestions for this work.
 L.X.G. wishes to thank National Council of Humanities, Sciences, and Technologies (CONAHCyT) for the support provided to her through the Cátedras CONAHCyT program and Project CBF2023-2024-268, Hadronic Physics at JLab: Deciphering the Internal Structure of Mesons and Baryons, from the 2023-2024 frontier science call.  A.R. acknowledges CIC-UMSNH for financial support under grant 18371.
 R.J.H.-P. are funded by CONAHCyT through Project No. 320856 (Paradigmas y Controversias de la Ciencia 2022) and Ciencia de frontera 2021-2042. 
  L.A. would like to acknowledge financial support by Ministerio Español de Ciencia e Innovación under grant No. PID2022-140440NB-C22; Junta de Andalucía under contract Nos. PAIDI FQM-370 and PCI+D+i under the title: “Tecnologías avanzadas para la exploración del universo y sus componentes" (Code AST22-0001).
 The authors are also funded by sistema Nacional de Investigadoras e investigadores from CONAHCyT.
\end{acknowledgements}
\appendix
\setcounter{equation}{0}
\renewcommand{\theequation}{\Alph{section}.\arabic{equation}}
\section{Euclidean Conventions}
\label{App:EM}
In our Euclidean formulation:
\begin{equation}
p\cdot q=\sum_{i=1}^4 p_i q_i\,;
\end{equation}
where
\begin{eqnarray}\nn
&&\{\gamma_\mu,\gamma_\nu\}=2\,\delta_{\mu\nu}\,;\;
\gamma_\mu^\dagger = \gamma_\mu\,;\;
\sigma_{\mu\nu}= \frac{i}{2}[\gamma_\mu,\gamma_\nu]\,; \; \\
&&{\rm tr}\,[\gamma_5\gamma_\mu\gamma_\nu\gamma_\rho\gamma_\sigma]=
-4\,\epsilon_{\mu\nu\rho\sigma}\,, \epsilon_{1234}= 1\,.
\end{eqnarray}
A positive energy spinor satisfies
\begin{equation}
\bar u(P,s)\, (i \gamma\cdot P + M) = 0 = (i\gamma\cdot P + M)\, u(P,s)\,,
\end{equation}
where $s=\pm$ is the spin label.  It is normalised as~:
\begin{equation}
\bar u(P,s) \, u(P,s) = 2 M \,,
\end{equation}
and may be expressed explicitly as~:
\begin{equation}
u(P,s) = \sqrt{M- i {\cal E}}
\left(
\begin{array}{c}
\chi_s\\
\displaystyle \frac{\vec{\sigma}\cdot \vec{P}}{M - i {\cal E}} \chi_s
\end{array}
\right)\,,
\end{equation}
with ${\cal E} = i \sqrt{\vec{P}^2 + M^2}$,
\begin{equation}
\chi_+ = \left( \begin{array}{c} 1 \\ 0  \end{array}\right)\,,\;
\chi_- = \left( \begin{array}{c} 0\\ 1  \end{array}\right)\,.
\end{equation}
For the free-particle spinor, $\bar u(P,s)= u(P,s)^\dagger \gamma_4$.
It can be used to construct a positive energy projection operator:
\begin{equation}
\label{Lplus} \Lambda_+(P):= \frac{1}{2 M}\,\sum_{s=\pm} \, u(P,s) \, \bar
u(P,s) = \frac{1}{2M} \left( -i \gamma\cdot P + M\right).
\end{equation}
A negative energy spinor satisfies
\begin{equation}
\bar v(P,s)\,(i\gamma\cdot P - M) = 0 = (i\gamma\cdot P - M) \, v(P,s)\,,
\end{equation}
and possesses properties and satisfies constraints obtained through obvious analogy
with $u(P,s)$. A charge-conjugated BSA is obtained via
\begin{equation}
\label{chargec}
\bar\Gamma(k;P) = C^\dagger \, \Gamma(-k;P)^{\rm T}\,C\,,
\end{equation}
where ``T'' denotes transposing  all matrix indices and
$C=\gamma_2\gamma_4$ is the charge conjugation matrix, $C^\dagger=-C$.  Moreover, we note that
\begin{equation}
C^\dagger \gamma_\mu^{\rm T} \, C = - \gamma_\mu\,, \; [C,\gamma_5] = 0\,.
\end{equation}
We employ a Rarita-Schwinger spinor to represent a covariant spin-$3/2$ field.  The positive energy
spinor is defined by the following equations:
\begin{equation}
\label{rarita}
(i \gamma\cdot P + M)\, u_\mu(P;r) = 0\,,
\gamma_\mu u_\mu(P;r) = 0\,,
P_\mu u_\mu(P;r) = 0,
\end{equation}
where $r=-3/2,-1/2,1/2,3/2$.  It is normalised as:
\begin{equation}
\bar u_{\mu}(P;r^\prime) \, u_\mu(P;r) = 2 M\,,
\end{equation}
and satisfies a completeness relation
\begin{equation}
\label{Deltacomplete}
\frac{1}{2 M}\sum_{r=-3/2}^{3/2} u_\mu(P;r)\,\bar u_\nu(P;r) =
\Lambda_+(P)\,R_{\mu\nu}\,,
\end{equation}
where
\begin{equation}
R_{\mu\nu} = \delta_{\mu\nu} \mbox{\boldmath $I$}_{\rm D} -\frac{1}{3} \gamma_\mu \gamma_\nu +
\frac{2}{3} \hat P_\mu \hat P_\nu \mbox{\boldmath $I$}_{\rm D} - i\frac{1}{3} [ \hat P_\mu
\gamma_\nu - \hat P_\nu \gamma_\mu]\,,
\label{Rde}\end{equation}
with $\hat P^2 = -1$. It is very useful in simplifying the FE for a positive energy decouplet state.
\section{ Contact interaction: features }
\label{CI-1}
The gap equation for fermions requires modelling the gluon propagator and the quark-gluon vertex. 
Here we shall recall and list these key characteristics of the CI~\cite{GutierrezGuerrero:2010md,Roberts:2010rn,Roberts:2011wy,Roberts:2011cf}~:
\begin{itemize}
\item  The gluon propagator is defined to be independent of any varying momentum scale:
\begin{eqnarray}
\label{eqn:contact_interaction}
g^{2}D_{\mu \nu}(k)&=&\frac{4\pi\alpha_{\mathrm{IR}}}{m_g^2}\delta_{\mu \nu} \equiv
\frac{1}{m_{G}^{2}}\delta_{\mu\nu},
\end{eqnarray}
\noindent where $m_g=500\,\MeV$ is a gluon mass scale generated dynamically in QCD~\cite{Boucaud:2011ug,Aguilar:2017dco,Binosi:2017rwj,Gao:2017uox}, and $\alpha_{\mathrm{IR}}$ can be interpreted as the interaction strength in the infrared~\cite{Binosi:2016nme,Deur:2016tte,Rodriguez-Quintero:2018wma}.
\item At  leading-order, the quark-gluon vertex is
\begin{equation}
\Gamma_{\nu}(p,q) =\gamma_{\nu}
\end{equation}
\item With this kernel the dressed-quark propagator for a quark of flavor $f$ becomes
\begin{eqnarray}
 \nn && S_f^{-1}(p) = \\
&&  i \gamma \cdot p + m_f +  \frac{16\pi}{3}\frac{\alpha_{\rm IR}}{m_G^2} \int\!\frac{d^4 q}{(2\pi)^4} \,
\gamma_{\mu} \, S_f(q) \, \gamma_{\mu}\,,\label{gap-1}
\end{eqnarray}
 where $m_f$ is the current-quark mass. The integral possesses quadratic and logarithmic divergences and we regularize them in a Poincar\'e covariant manner. The solution of this equation is~: 
\begin{equation}
\label{genS}
S_f^{-1}(p) = i \gamma\cdot p + M_f\,,
\end{equation}
 where $M_f$ is, in general a mass function running with a momentum scale, but within the CI it 
 is a constant dressed mass.
\item $M_f$ is determined by
\begin{equation}
M_f = m_f + M_f\frac{4\alpha_{\rm IR}}{3\pi m_G^2}\,\,{\cal C}^{\rm iu}(M_f^2)\,,
\label{gapactual}
\end{equation}
where 
\bea
\hspace{0.75 cm}
\nn {\cal C}^{\rm iu}(\sigma)/\sigma = \overline{\cal C}^{\rm iu}(\sigma) = \Gamma(-1,\sigma \tau_{\rm UV}^2) - \Gamma(-1,\sigma \tau_{\rm IR}^2),\\
\label{Ciu}
\eea
 with $\Gamma(\alpha,y)$ being the incomplete gamma-function and $\tau_{\rm IR, UV}$ are respectively, infrared and ultraviolet regulators. A nonzero value for  $\tau_{\mathrm{IR}}\equiv 1/\Lambda_{\mathrm{IR}}$ implements 
confinement~\cite{Roberts:2007ji}. Since the CI is a nonrenormalizable theory, 
$\tau_{\mathrm{UV}}\equiv 1/\Lambda_{\mathrm{UV}}$ becomes part of the model and therefore sets the scale for
all dimensional quantities. 
 \end{itemize}
 In this work we report results  using the parameter set in Table~\ref{parameters1}. 
 \begin{table}[htbp]
 \caption{\label{parameters1} Dimensionless coupling constant $\hat{\alpha}=\hat{\alpha}_{\mathrm IR}\Lambda_{\mathrm UV}^{2}$,
  where $\rmh=\alpha_{\mathrm {IR}}/m_{g}^{2}$, for the CI, extracted from a
  best-fit to data, as explained in Ref.~\cite{Raya:2017ggu}. Fixed parameters are $m_{g}~=~0.50~\GeV$ and
  $\Lambda_{\mathrm {IR}}=0.24\,\GeV$.}
\begin{center}
\begin{tabular}{@{\extracolsep{0.5 cm}}lcc}
\hline \hline
 quarks & $\hat{\alpha}_{\mathrm {IR}}\;[\GeV^{-2}]$ & $\Lambda_{\mathrm {UV}}\;[\GeV] $  \\
  \rule{0ex}{3.5ex}
$\tu$, $\td$, $\ts$ & 1.14 & 0.91\\ 
 \rule{0ex}{3.5ex}
$\tc,\td,\ts$ & 0.38 & 1.32 \\
 \rule{0ex}{3.5ex}
$\tc$     & 0.09 & 2.31 \\
 \rule{0ex}{3.5ex}
$\tb,\tu,\ts$ & 0.07 & 2.52 \\
 \rule{0ex}{3.5ex}
$\tb,\tc$   & 0.02 & 4.13 \\
 \rule{0ex}{3.5ex}
$\tb$     & 0.007 & 6.56\\
\hline \hline
\end{tabular}
\end{center}
\end{table}
 %
 %
%
%
 Table~\ref{table-M} presents the values of $\tu$, $\ts$, $\tc$ and $\tb$ dressed quark masses computed from Eq.~(\ref{gapactual}).
%
\begin{table}[htbp]
\caption{\label{table-M}
Computed dressed-quark masses (in \GeV), required as input for the BS equation and FE.}
\vspace{-0cm}
\begin{tabular}{@{\extracolsep{0.2 cm}}cccc}
\hline 
\hline
 $m_{\tu}=0.007$ &$m_{\ts}=0.17$ & $m_{\tc}=1.08$ & $m_{\tb}=3.92$   \\
 $\hspace{-.2cm} M_{\tu}=0.36$ & $ \hspace{0.1cm} M_{\ts}=0.53$\; &  $ \hspace{0.0 cm} M_{\tc}=1.52$ & $ \hspace{0.0cm} M_{\tb}=4.68$  \\
 \hline
 \hline
\end{tabular}
\end{table}
The simplicity of the CI allows one to readily compute hadronic observables, such as masses, decay constants, charge radii and form factors. 
\section{Masses of mesons and diquarks containing c and b quarks}
\label{BS-s}
\begin{figure}[ht]
\vspace{-5.0cm}
       \centerline{
\includegraphics[scale=0.45,angle=0]{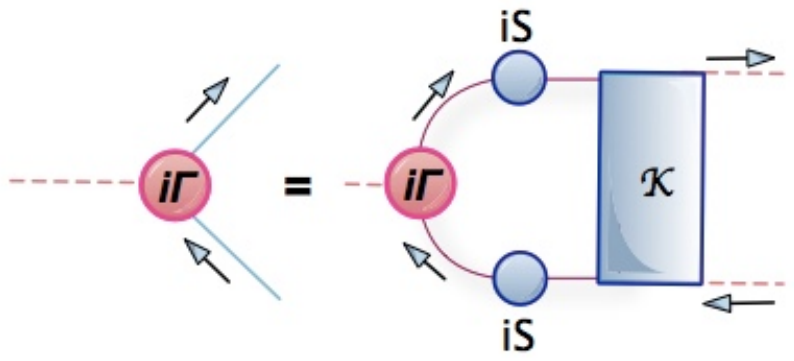}}
       \vspace{-5.0cm}
       \caption{\label{fig:BSEfig}This diagram represents the BS equation. Blue (solid) circles represent dressed propagators $S$, red (solid) circle is the meson BSA $\Gamma$ and the blue (solid) rectangle is the dressed-quark-antiquark scattering kernel $K$.}
\end{figure}

The bound-state problem for hadrons characterized by two valence-fermions may be studied using the
homogeneous BS equation in the \fig{fig:BSEfig}.
\\
 The corresponding equation is~\cite{Salpeter:1951sz}
\begin{equation}
[\Gamma(k;P)]_{tu} = \int \! \frac{d^4q}{(2\pi)^4} [\chi(q;P)]_{sr} K_{tu}^{rs}(q,k;P)\,,
\label{genbse}
\end{equation}
where $\Gamma$ is the bound-state's BSA; $\chi(q;P) = S(q+P)\Gamma S(q)$ is the BS wave-function; $r,s,t,u$ represent colour, flavor and spinor indices; and $K$ is the relevant fermion-fermion scattering kernel.  This equation possesses solutions on that discrete set of $P^2$-values for which bound-states exist.\\
A general decomposition of the bound-state's BSA for scalar and axial-vector diquarks composed of quarks $\tq$ and $\tqu$ in the CI model has the form~:
\bea
\label{BSA-diquarks}
\nn \Gamma_{[\tq\tqu]}^{0^+}&=&i\gamma_5E_{[\tq\tqu]_{0^+}} +\frac{1}{2M_R}\gamma_5 \gamma\cdot P F_{[\tq\tqu]_{0^+}}\;,\\
\Gamma_{\{\tq\tqu\},\mu}^{1^+}&=&\gamma_{\mu}^{T}E_{\{\tq\tqu\}_{1^+}} +\frac{1}{2M_R}\sigma_{\mu\nu}P_{\nu}F_{\{\tq\tqu\}_{1^+}}\;,
\eea
where $M_R = M_{\tq} M_{\tqu}/[M_{\tq} + M_{\tqu}]$ and $M_{\tq}$ and $M_{\tqu}$ are the quark masses. The amplitudes in \eqn{BSA-diquarks} are crucial for determining the masses listed in Table~\ref{fbod}.
\setcounter{equation}{0}
\section{Kernel in FE}\label{app:Fad}
In this section, we provide the explicit expressions for the elements of the matrix in the \eqn{B-matrix},
\begin{align}
\nn \mathcal{M}^{11}&= t^{\tq T}t^{[\tq\tqu]}t^{[\tq\tqu]T}t^{\tq}\\
\nn &\times\{\g^{0^+}_{[\tq\tqu]}(l_{\tq\tqu}) S_{\tqu}^T\overline{\g}^{0^+}_{[\tq\tqu]}(-k_{\tq\tqu})S_{\tq}(l_{\tq})\D^{0^+}_{[\tq\tqu]}(l_{\tq\tqu})\}\;,\\
\nn \mathcal{M}^{12}_{\nu}&= t^{\tq T}t^{\{\tq\tq\}}t^{[\tq\tqu]T}t^{\tqu} \\
\nn &\times\{\g^{1^+}_{\{\tq\tq\},\mu}(l_{\tq\tq}) S_{\tq}^T\overline{\g}^{0^+}_{[\tq\tqu]}(-k_{\tq\tqu})S_{\tqu}(l_{\tqu})\D^{1^+}_{\{\tq\tq\},\mu\nu}(l_{\tq\tq})\}\;,\\
\nn \mathcal{M}^{13}_\nu&= t^{\tq T}t^{\{\tq\tqu\}}t^{[\tq\tqu]T}t^{\tq}\\
 \nn & \times \{\g^{1^+}_{\{\tq\tqu\},\mu}(l_{\tq\tqu}) S_{\tqu}^T\overline{\g}^{0^+}_{[\tq\tqu]}(-k_{\tq\tqu})S_{\tq}(l_{\tq})\D^{1^+}_{\{\tq\tqu\},\mu\nu}(l_{\tq\tqu})\}\;,\\
\nn \mathcal{M}^{14}_\nu&= t^{\tq T}t^{\{\tq\tq\}}t^{[\tq\tqu]T}t^{\tqu}\\
\nn & \times \g^{1^+}_{\{\tq\tq\},\mu}(l_{\tq\tq}) S_{\tq}^T\overline{\g}^{0^+}_{[\tq\tqu]}(-k_{\tq\tqu})S_{\tqu}(l_{\tqu})\D^{1^+}_{\{\tq\tq\},\mu\nu}(l_{\tq\tq})\}\;,\\
\nn\mathcal{M}^{15}_\nu&= t^{\tq T}t^{\{\tq\tqu\}}t^{[\tq\tqu]T}t^{\tq}\\
\nn \times & \{\g^{1^+}_{\{\tq\tqu\},\mu}(l_{\tq\tqu}) S_{\tqu}^T\overline{\g}^{0^+}_{[\tq\tqu]}(-k_{\tq\tqu})S_{\tq}(l_{\tq})\D^{1^+}_{\{\tq\tqu\},\mu\nu}(l_{\tq\tqu})\}\;,\\
\nn \mathcal{M}^{21}_\mu&=t^{\tqu T}t^{[\tq\tqu]}t^{\{\tq\tq\}T}t^{\tq}\\
\nn\times &\{\{\g^{0^+}_{[\tq\tqu]}(l_{\tq\tqu}) S_{\tq}^T\overline{\g}^{1^+}_{\{\tq\tq\},\mu}(-k_{\tq\tq})S_{\tq}(l_{\tq})\D^{0^+}_{[\tq\tqu]}(l_{\tq\tqu})\}\;,\\
\nn  \mathcal{M}^{22}_{\mu\nu} &=t^{\tqu T}t^{\{\tq\tq\}}t^{\{\tq\tq\}T}t^{\tqu}
\\ \nn\times &
\nn \{\g^{1^+}_{\{\tq\tq\},\rho}(l_{\tq\tq}) S_{\tqu}^T\overline{\g}^{1^+}_{\{\tq\tq\},\mu}(-k_{\tq\tq})S_{\tqu}(l_{\tqu})\D^{1^+}_{\{\tq\tq\},\rho\nu}(l_{\tq\tq})\}\;,
\end{align}
\begin{align}
\nn \mathcal{M}^{23}_{\mu\nu}&=t^{\tqu T}t^{\{\tq\tqu\}}t^{\{\tq\tq\}T}t^{\tq}\\
\nn \times &\{\g^{1^+}_{\{\tq\tqu\},\rho}(l_{\tq\tqu}) S_{\tq}^T\overline{\g}^{1^+}_{\{\tq\tq\},\mu}(-k_{\tq\tq})S_{\tq}(l_{\tq})\D^{1^+}_{\{\tq\tqu\},\rho\nu}(l_{\tq\tqu})\}\;,\\
\nn \mathcal{M}^{24}_{\mu\nu} &=t^{\tqu T}t^{\{\tq\tq\}}t^{\{\tq\tq\}T}t^{\tqu}\\
\nn \times & \{\g^{1^+}_{\{\tq\tq\},\rho}(l_{\tq\tq}) S_{\tqu}^T\overline{\g}^{1^+}_{\{\tq\tq\},\mu}(-k_{\tq\tq})S_{\tqu}(l_{\tqu})\D^{1^+}_{\{\tq\tq\},\rho\nu}(l_{\tq\tq})\}\;,\\
\nn \mathcal{M}^{25}_{\mu\nu} &=t^{\tqu T}t^{\{\tq\tqu\}}t^{\{\tq\tq\}T}t^{\tq}\\
\nn\times &\{\g^{1^+}_{\{\tq\tqu\},\rho}(l_{\tq\tqu}) S_{\tq}^T\overline{\g}^{1^+}_{\{\tq\tq\},\mu}(-k_{\tq\tq})S_{\tq}(l_{\tq})\D^{1^+}_{\{\tq\tqu\},\rho\nu}(l_{\tq\tqu})\}\;,\\
\nn \mathcal{M}^{31}_{\mu}\ &=t^{\tq T}t^{[\tq\tqu]}t^{\{\tq\tqu\}T}t^{\tq}\\
\times &
\nn \{\g^{0^+}_{[\tq\tqu]}(l_{\tq\tqu}) S_{\tqu}^T\overline{\g}^{1^+}_{\{\tq\tqu\},\mu}(-k_{\tq\tqu})S_{\tq}(l_{\tq})\D^{0^+}_{[\tq\tqu]}(l_{\tq\tqu})\}\;,\\
\nn \mathcal{M}^{32}_{\mu\nu}&=t^{\tq T}t^{\{\tq\tq\}}t^{\{\tq\tq\}T}t^{\tqu}\\
\nn \times
&\{\g^{1^+}_{\{\tq\tq\},\rho}(l_{\tq\tq}) S_{\tq}^T\overline{\g}^{1^+}_{\{\tq\tqu\},\mu}(-k_{\tq\tqu})S_{\tqu}(l_{\tqu})\D^{0^+}_{\{\tq\tq\},\rho\nu}(l_{\tq\tq})\}\;,\\
\nn \mathcal{M}^{33}_{\mu\nu}&=t^{\tq T}t^{\{\tq\tqu\}}t^{\{\tq\tqu\}T}t^{\tq}\\
\nn \times & \{\g^{1^+}_{\{\tq\tqu\},\rho}(l_{\tq\tqu}) S_{\tq}^T\overline{\g}^{1^+}_{\{\tq\tqu\},\mu}(-k_{\tq\tqu})S_{\tq}(l_{\tq})\D^{0^+}_{\{\tq\tqu\},\rho\nu}(l_{\tq\tqu})\}\;,\\
\nn \mathcal{M}^{34}_{\mu\nu}&=t^{\tq T}t^{\{\tq\tq\}}t^{\{\tq\tq\}T}t^{\tqu}\\
\nn \times & \{\g^{1^+}_{\{\tq\tq\},\rho}(l_{\tq\tq}) S_{\tq}^T\overline{\g}^{1^+}_{\{\tq\tqu\},\mu}(-k_{\tq\tqu})S_{\tqu}(l_{\tqu})\D^{0^+}_{\{\tq\tq\},\rho\nu}(l_{\tq\tq})\}\;,\\
\nn  \mathcal{M}^{35}_{\mu\nu}&=t^{\tq T}t^{\{\tq\tqu\}}t^{\{\tq\tqu\}T}t^{\tq}\\
\nn \times & \{\g^{1^+}_{\{\tq\tqu\},\rho}(l_{\tq\tqu}) S_{\tq}^T\overline{\g}^{1^+}_{\{\tq\tqu\},\mu}(-k_{\tq\tqu})S_{\tq}(l_{\tq})\D^{0^+}_{\{\tq\tqu\},\rho\nu}(l_{\tq\tqu})\}\;,\\
\nn \mathcal{M}^{41}_\mu&=t^{\tqu T}t^{[\tq\tqu]}t^{\{\tq\tq\}T}t^{\tq}\\
\nn \times &\{\{\g^{0^+}_{[\tq\tqu]}(l_{\tq\tqu}) S_{\tq}^T\overline{\g}^{1^+}_{\{\tq\tq\},\mu}(-k_{\tq\tq})S_{\tq}(l_{\tq})\D^{0^+}_{[\tq\tqu]}(l_{\tq\tqu})\}\;,\\
\nn  \mathcal{M}^{42}_{\mu\nu}&=t^{\tqu T}t^{\{\tq\tq\}}t^{\{\tq\tq\}T}t^{\tqu}\\
\nn \times & \{\g^{1^+}_{\{\tq\tq\},\rho}(l_{\tq\tq}) S_{\tqu}^T\overline{\g}^{1^+}_{\{\tq\tq\},\mu}(-k_{\tq\tq})S_{\tqu}(l_{\tqu})\D^{1^+}_{\{\tq\tq\},\rho\nu}(l_{\tq\tq})\}\;,\\
\nn \mathcal{M}^{43}_{\mu\nu}&=t^{\tqu T}t^{\{\tq\tqu\}}t^{\{\tq\tq\}T}t^{\tq}\\
\nn & \times\{\g^{1^+}_{\{\tq\tqu\},\rho}(l_{\tq\tq}) S_{\tq}^T\overline{\g}^{1^+}_{\{\tq\tq\},\mu}(-k_{\tq\tq})S_{\tq}(l_{\tq})\D^{1^+}_{\{\tq\tqu\},\rho\nu}(l_{\tq\tqu})\}\;,\\
\nn  \mathcal{M}^{44}_{\mu\nu} &=t^{\tqu T}t^{\{\tq\tq\}}t^{\{\tq\tq\}T}t^{\tqu} \\
\nn \times & \{\g^{1^+}_{\{\tq\tq\},\rho}(l_{\tq\tq}) S_{\tqu}^T\overline{\g}^{1^+}_{\{\tq\tq\},\mu}(-k_{\tq\tq})S_{\tqu}(l_{\tqu})\D^{1^+}_{\{\tq\tq\},\rho\nu}(l_{\tq\tq})\}\;,\\
\nn \mathcal{M}^{45}_{\mu\nu}&=t^{\tqu T}t^{\{\tq\tqu\}}t^{\{\tq\tq\}T}t^{\tq} \\
\nn
\times & \{\g^{1^+}_{\{\tq\tqu\},\rho}(l_{\tq\tqu}) S_{\tq}^T\overline{\g}^{1^+}_{\{\tq\tq\},\mu}(-k_{\tq\tq})S_{\tq}(l_{\tqu})\D^{1^+}_{\{\tq\tqu\},\rho\nu}(l_{\tq\tqu})\}\;,\\
\nn  \mathcal{M}^{51}_{\mu}&=t^{\tq T}t^{[\tq\tqu]}t^{\{\tq\tqu\}T}t^{\tq}\\
\nn \times & \{\g^{0^+}_{[\tq\tqu]}(l_{\tq\tqu}) S_{\tqu}^T\overline{\g}^{1^+}_{\{\tq\tqu\},\mu}(-k_{\tq\tqu})S_{\tq}(l_{\tq})\D^{0^+}_{[\tq\tqu]}(l_{\tq\tqu})\}\;,\\
\nn \mathcal{M}^{52}_{\mu\nu}&=t^{\tq T}t^{\{\tq\tq\}}t^{\{\tq\tq\}T}t^{\tqu}\\
\nn \times & \{\g^{1^+}_{\{\tq\tq\},\rho}(l_{\tq\tq}) S_{\tq}^T\overline{\g}^{1^+}_{\{\tq\tqu\},\mu}(-k_{\tq\tqu})S_{\tqu}(l_{\tqu})\D^{0^+}_{\{\tq\tq\},\rho\nu}(l_{\tq\tq})\}\;,\\
\nn \mathcal{M}^{53}_{\mu\nu}&=t^{\tq T}t^{\{\tq\tqu\}}t^{\{\tq\tqu\}T}t^{\tq}\\
\nn \times &\{\g^{1^+}_{\{\tq\tqu\},\rho}(l_{\tq\tqu}) S_{\tq}^T\overline{\g}^{1^+}_{\{\tq\tqu\},\mu}(-k_{\tq\tqu})S_{\tq}(l_{\tq})\D^{0^+}_{\{\tq\tqu\},\rho\nu}(l_{\tq\tqu})\}\;,\\
\nn \mathcal{M}^{54}_{\mu\nu}&=t^{\tq T}t^{\{\tq\tq\}}t^{\{\tq\tq\}T}t^{\tqu}
 \\ \nn \times & \{\g^{1^+}_{\{\tq\tq\},\rho}(l_{\tq\tq}) S_{\tq}^T\overline{\g}^{1^+}_{\{\tq\tqu\},\mu}(-k_{\tq\tqu})S_{\tqu}(l_{\tqu})\D^{0^+}_{\{\tq\tq\},\rho\nu}(l_{\tq\tq})\}\;,
\end{align}
\begin{align}
\nn \mathcal{M}^{55}_{\mu\nu}&=t^{\tq T}t^{\{\tq\tqu\}}t^{\{\tq\tqu\}T}t^{\tq}
\\ \nn \times &\{\g^{1^+}_{\{\tq\tqu\},\rho}(l_{\tq\tqu}) S_{\tq}^T\overline{\g}^{1^+}_{\{\tq\tqu\},\mu}(-k_{\tq\tqu})S_{\tq}(l_{\tq})\D^{0^+}_{\{\tq\tqu\},\rho\nu}(l_{\tq\tqu})\}\;.
\end{align}
 $\Delta_{[\tq\tqu]}^{0^+}(K)$ and $\Delta_{\{\tq\tqu\},\mu\nu}^{1^+}(K)$, are standard
 propagators for scalar and vector diquarks.
 \begin{eqnarray}
&&\Delta_{[\tq\tqu]}^{0^+}(K)=\frac{1}{K^2\!+\!m_{\tq\tqu}^2},\\
&&\Delta_{\{\tq\tqu\},\mu\nu}^{1^+}(K)\!=\!\frac{1}{K^2+m_{\tq\tqu}^2}\left(\!\delta_{\mu\nu}\!+\!\frac{K_\mu K_\nu}{m_{\tq\tqu}^2}\!\right).
\end{eqnarray}
With these expressions, the calculation of the baryonic masses with spin $1/2$ in Section \ref{Baryons} is straightforward.
\section{Flavor Diquarks}
\label{app:Fla}
We define the following set of flavor column matrices,
\begin{equation}\nn
t^{\tu}=\begin{pmatrix} 1  \\ 0  \\ 0 \\ 0 \\0 \end{pmatrix},\;\;\;\;
t^{\td}=\begin{pmatrix} 0  \\ 1  \\ 0  \\ 0\\0  \end{pmatrix},\;\;\;\;
t^{\ts}=\begin{pmatrix} 0  \\ 0  \\ 1 \\ 0 \\0  \end{pmatrix},\;\;\;\;
\end{equation}
\begin{equation}
t^{\tc}=\begin{pmatrix} 0  \\ 0  \\ 0 \\ 1\\0  \end{pmatrix},\;\;\;\;
t^{\tb}=\begin{pmatrix} 0  \\ 0  \\ 0 \\ 0\\1  \end{pmatrix},\;\;\;\;
\end{equation}
and
\bea \nn
t^f=\left[\begin{matrix} \nn
t^{\tq T}t^{\{\tqu \tq\}}t^{\{\tqu\tq\}T}t^{\tq} &&t^{\tq T}t^{\{\tq\tq\}}t^{\{\tqu\tq\}T}t^{\tqu}\\ \\
t^{\tqu T}t^{\{\tqu\tq\}}t^{\{\tq\tq\}T}t^{\tq} && t^{\tqu T}t^{\{\tq\tq\}}t^{\{\tq\tq\}T}t^{\tqu}
\end{matrix}\right] \,.
\eea
The flavor matrices for the diquarks are
\begin{equation}\nonumber
\begin{array}{cc}
\bf{t^{[\tu\td]}=\begin{pmatrix} 0 & 1 & 0 & 0 & 0 \\ -1 & 0 & 0  & 0& 0 \\ 0 &  0 & 0  & 0 & 0 \\ 0 &  0 & 0  & 0 & 0\\  0 & 0 & 0 & 0 & 0 \end{pmatrix}} 
&
\bf{t^{[\tu\ts]}=\begin{pmatrix} 0 & 0 & 1& 0 & 0  \\ 0 & 0 & 0& 0& 0 \\ -1 &  0 & 0& 0 & 0\\ 0 &  0 & 0  & 0& 0\\  0 & 0 & 0 & 0 & 0 \end{pmatrix}}
\\[12ex]
\bf{t^{[\td\ts]}=\begin{pmatrix} 0 & 0 & 0 & 0 & 0 \\ 0 & 0 & 1& 0 & 0  \\ 0 &  -1 & 0 & 0& 0   \\ 0 &  0 & 0  & 0& 0 \\  0 & 0 & 0 & 0 & 0 \end{pmatrix}}
&
\bf{t^{[\tu\tc]}=\begin{pmatrix} 0 & 0 & 0 & 1& 0  \\ 0 & 0 & 0& 0& 0   \\ 0 &  0 & 0 & 0& 0   \\ -1 &  0 & 0  & 0& 0\\  0 & 0 & 0 & 0 & 0 \end{pmatrix}},
\end{array}
 \end{equation}
\begin{equation}\nonumber
\begin{array}{cc}
\bf{t^{\{\tu\tu\}}=\begin{pmatrix} \sqrt{2} & 0 & 0& 0 &0\\ 0 & 0 & 0& 0&0 \\ 0 &  0 & 0 & 0&0 \\ 0 &  0 & 0 & 0&0\\ 0 & 0 & 0& 0& 0  \end{pmatrix}}
&
\bf{t^{\{\tu\td\}}=\begin{pmatrix} 0 & 1 & 0 & 0&0\\ 1 & 0 & 0& 0&0 \\ 0 &  0 & 0& 0&0 \\ 0 &  0 & 0 & 0 &0\\ 0 & 0 & 0& 0& 0  \end{pmatrix}},\\[10ex]
\bf{t^{\{\tu\ts\}}=\begin{pmatrix} 0 & 0 & 1& 0&0  \\ 0 & 0 & 0& 0&0  \\ 1 &  0 & 0& 0 &0 \\ 0 &  0 & 0 & 0&0 \\ 0 & 0 & 0& 0& 0   \end{pmatrix}}
&
t^{\{\tu\tc\}}=\begin{pmatrix} 0 & 0 & 0& 1&0 \\ 0 & 0 & 0 & 0&0\\ 0 &  0 & 0 & 0&0\\  1 &  0 & 0 & 0&0\\ 0 & 0 & 0& 0& 0  \end{pmatrix},
\\[10ex]
\bf{t^{\{\td\td\}}=\begin{pmatrix} 0 & 0 & 0 & 0 & 0 \\ 0 &  \sqrt{2} & 0& 0 & 0 \\ 0 &  0 & 0& 0& 0  \\ 0 & 0 & 0 & 0& 0\\ 0 & 0 & 0& 0& 0  \end{pmatrix}}
&
\bf{t^{\{\td\ts\}}}=\begin{pmatrix} 0 & 0 & 0& 0 & 0 \\ 0 & 0 & 1 & 0 & 0\\ 0 &  1 & 0 & 0 & 0\\ 0 & 0 & 0 & 0& 0\\ 0 & 0 & 0& 0& 0   \end{pmatrix},
\\ [10ex]
\bf{t^{\{\ts\ts\}}=\begin{pmatrix} 0 & 0 & 0& 0& 0 \\ 0 & 0 & 0& 0& 0 \\ 0 &  0 & \sqrt{2} & 0 & 0 \\ 0 & 0 & 0 & 0 & 0\\ 0 & 0 & 0& 0& 0  \end{pmatrix}}, 
&
t^{\{\tc\tc\}}=\begin{pmatrix} 0 & 0 & 0& 0 & 0 \\ 0 & 0 & 0 & 0 & 0\\ 0 &  0 & 0 & 0 & 0\\  0 &  0 & 0 &  \sqrt{2} & 0\\ 0 & 0 & 0& 0& 0   \end{pmatrix}\,,
\\[10ex]
\bf{t^{[\td\tc]}}=\begin{pmatrix} 0 & 0 & 0& 0& 0 \\ 0 & 0 & 0 & 1& 0\\ 0 &  0 & 0 & 0& 0\\  0 &  -1& 0 & 0& 0 \\ 0 & 0 & 0& 0& 0    \end{pmatrix},
&
\bf{t^{[\ts\tc]}}=\begin{pmatrix} 0 & 0 & 0& 0 & 0 \\ 0 & 0 & 0 & 0& 0\\ 0 &  0 & 0 & 1& 0\\  0 &  0& -1 & 0& 0 \\ 0 & 0 & 0& 0& 0    \end{pmatrix},
%
\\[10ex]
\bf{t^{\{\td\tc\}}}=\begin{pmatrix} 0 & 0 & 0& 0 & 0\\ 0 & 0 & 0 & 1& 0\\ 0 &  0 & 0 & 0& 0\\  0 &  1& 0 & 0& 0 \\ 0 & 0 & 0& 0& 0   \end{pmatrix},
&
\bf{t^{\{\ts\tc\}}}=\begin{pmatrix} 0 & 0 & 0& 0 & 0 \\ 0 & 0 & 0 & 0 & 0\\ 0 &  0 & 0 & 1 & 0\\  0 &  0& 1 & 0 & 0 \\ 0 & 0 & 0& 0& 0   \end{pmatrix},\\[10ex]
\bf{t^{[\tb\tu]}}=\begin{pmatrix} 0 & 0 & 0& 0& -1\\ 0 & 0 & 0 & 0& 0\\ 0 &  0 & 0 & 0& 0\\  0 &  0& 0 & 0& 0 \\ 1 & 0 & 0& 0& 0    \end{pmatrix},
&
\bf{t^{[\tb\td]}}=\begin{pmatrix} 0 & 0 & 0& 0 & 0 \\ 0 & 0 & 0 & 0& -1\\ 0 &  0 & 0 & 0& 0\\  0 &  0& 0 & 0& 0 \\ 0 & 1 & 0& 0& 0    \end{pmatrix},
\\[10ex]
\bf{t^{[\tb\tc]}}=\begin{pmatrix} 0 & 0 & 0& 0 & 0 \\ 0 & 0 & 0 & 0& 0\\ 0 &  0 & 0 & 0& 0\\  0 &  0& 0 & 0& -1 \\ 0 & 0 & 0& 1& 0    \end{pmatrix},
&
\bf{t^{[\tb\ts]}}=\begin{pmatrix} 0 & 0 & 0& 0 & 0 \\ 0 & 0 & 0 & 0& 0\\ 0 &  0 & 0 & 0& -1\\  0 &  0& 0 & 0& 0 \\ 0 & 0 & 1& 0& 0    \end{pmatrix},\;\;\;\;
\\[10ex]
\bf{t^{\{\tb\tb\}}}=\begin{pmatrix} 0 & 0 & 0& 0& 0\\ 0 & 0 & 0 & 0& 0\\ 0 &  0 & 0 & 0& 0\\  0 &  0& 0 & 0& 0 \\ 0 & 0 & 0& 0&  \sqrt{2}    \end{pmatrix}
&
\bf{t^{\{\tb\tu\}}}=\begin{pmatrix} 0 & 0 & 0& 0& 1\\ 0 & 0 & 0 & 0& 0\\ 0 &  0 & 0 & 0& 0\\  0 &  0& 0 & 0& 0 \\ 1 & 0 & 0& 0& 0 \end{pmatrix},\;\;\;\;
\end{array}
\end{equation}
\begin{equation}\nonumber
\begin{array}{cc}
\bf{t^{\{\tb\td\}}}=\begin{pmatrix} 0 & 0 & 0& 0& 0\\ 0 & 0 & 0 & 0& 1\\ 0 &  0 & 0 & 0& 0\\  0 &  0& 0 & 0& 0 \\ 0 & 1 & 0& 0& 0   \end{pmatrix}
&
\bf{t^{\{\tb\tc\}}}=\begin{pmatrix} 0 & 0 & 0& 0& 0\\ 0 & 0 & 0 & 0& 0\\ 0 &  0 & 0 & 0& 0\\  0 &  0& 0 & 0& 1 \\ 0 & 0 & 0& 1& 0   \end{pmatrix},
\\[10ex]
\bf{t^{\{\tb\ts\}}}=\begin{pmatrix} 0 & 0 & 0& 0& 0\\ 0 & 0 & 0 & 0& 0\\ 0 &  0 & 0 & 0& 1\\  0 &  0& 0 & 0& 0 \\ 0 & 0 & 1& 0& 0   
&
 \end{pmatrix}
 \end{array}
\end{equation}
\bibliography{ccc-a}

\end{document}